\documentstyle[12pt,epsfig,color]{article}
\def\baselinestretch{1.3}
\advance\textheight by \topskip
\newcommand{\comment}[1]{}

\def\beq{\begin{equation}}
\def\eeq{\end{equation}}
\def\beqn{\begin{eqnarray}}
\def\eeqn{\end{eqnarray}}

\begin{document}
 \tolerance=100000
 \topmargin -0.1in
\headsep 30pt
\footskip 40pt
\oddsidemargin 12pt
\evensidemargin -16pt
\textheight 8.5in
\textwidth 6.5in
\parindent 20pt
 
\def\baselinestretch{1.5}
\newcommand{\newc}{\newcommand}
\def\preprint{{preprint}}
\def\Ord{\lower .7ex\hbox{$\;\stackrel{\textstyle <}{\sim}\;$}}
\def\OOrd{\lower .7ex\hbox{$\;\stackrel{\textstyle >}{\sim}\;$}}
\def\cO#1{{\cal{O}}\left(#1\right)}
\newc{\order}{{\cal O}}
\def\lag             {{\cal L}}
\def\Lag             {{\cal L}}
\def\lum             {{\cal L}}
\def\R               {{\cal R}}
\def\Rsq             {{\cal R}^{\sq}}
\def\Rst             {{\cal R}^{\st}}
\def\Rsb             {{\cal R}^{\sb}}
\def\M               {{\cal M}}
\def\Oas             {{\cal O}(\alpha_{s})}
\def\Vcal            {{\cal V}}
\def\Wcal            {{\cal W}}
\newc{\be}{\begin{equation}}
\newc{\ee}{\end{equation}}
\newc{\br}{\begin{eqnarray}}
\newc{\er}{\end{eqnarray}}
\newc{\ba}{\begin{array}}
\newc{\ea}{\end{array}}
\newc{\bi}{\begin{itemize}}
\newc{\ei}{\end{itemize}}
\newc{\bn}{\begin{enumerate}}
\newc{\en}{\end{enumerate}}
\newc{\bc}{\begin{center}}
\newc{\ec}{\end{center}}
\newc{\ul}{\underline}
\newc{\ol}{\overline}
\newc{\ra}{\rightarrow}
\newc{\lra}{\longrightarrow}
\newc{\wt}{\widetilde}
\newc{\til}{\tilde}
\def\kr              {^{\dagger}}
\newc{\wh}{\widehat}
\newc{\ti}{\times}
\newc{\Dir}{\kern -6.4pt\Big{/}}
\newc{\Dirin}{\kern -10.4pt\Big{/}\kern 4.4pt}
\newc{\DDir}{\kern -10.6pt\Big{/}}
\newc{\DGir}{\kern -6.0pt\Big{/}}
\newc{\sig}{\sigma}
\newc{\sigmalstop}{\sig_{\lstoppair}}
\newc{\Sig}{\Sigma}  
\newc{\del}{\delta}
\newc{\Del}{\Delta}
\newc{\lam}{\lambda}
\newc{\Lam}{\Lambda}
\newc{\gam}{\gamma}
\newc{\Gam}{\Gamma}
\newc{\eps}{\epsilon}
\newc{\Eps}{\Epsilon}
\newc{\kap}{\kappa}
\newc{\Kap}{\Kappa}
\newc{\modulus}[1]{\left| #1 \right|}
\newc{\eq}[1]{(\ref{eq:#1})}
\newc{\eqs}[2]{(\ref{eq:#1},\ref{eq:#2})}
\newc{\etal}{{\it et al.}\ }
\newc{\ibid}{{\it ibid}.}
\newc{\ibidem}{{\it ibidem}.}
\newc{\eg}{{\it e.g.}\ }
\newc{\ie}{{\it i.e.}\ }
\def \viz{\emph{viz.}}
\def \etc{\emph{etc. }}
\newc{\nonum}{\nonumber}
\newc{\lab}[1]{\label{eq:#1}}
\newc{\dpr}[2]{({#1}\cdot{#2})}
\newc{\lt}{\stackrel{<}}
\newc{\gt}{\stackrel{>}}
\newc{\lsimeq}{\stackrel{<}{\sim}}
\newc{\gsimeq}{\stackrel{>}{\sim}}
\def\lsim{\buildrel{\scriptscriptstyle <}\over{\scriptscriptstyle\sim}}
\def\gsim{\buildrel{\scriptscriptstyle >}\over{\scriptscriptstyle\sim}}
\def\lapp{\mathrel{\rlap{\raise.5ex\hbox{$<$}}
                    {\lower.5ex\hbox{$\sim$}}}}
\def\gapp{\mathrel{\rlap{\raise.5ex\hbox{$>$}}
                    {\lower.5ex\hbox{$\sim$}}}}
\newc{\half}{\frac{1}{2}}
\newcommand {\nnc}        {{\overline{\mathrm N}_{95}}}
\newcommand {\dm}         {\Delta m}
\newcommand {\dM}         {\Delta M}
\def\bra{\langle}
\def\ket{\rangle}
\def\cO#1{{\cal{O}}\left(#1\right)}
\def \DM{{\Delta{m}}}
\newc{\bQ}{\ol{Q}}
\newc{\dota}{\dot{\alpha }}
\newc{\dotb}{\dot{\beta }}
\newc{\dotd}{\dot{\delta }}
\newc{\nindnt}{\noindent}

\newcommand{\medf}[2] {{\footnotesize{\frac{#1}{#2}} }}
\newcommand{\smaf}[2] {{\textstyle \frac{#1}{#2} }}
\def\onesq            {{\textstyle \frac{1}{\sqrt{2}} }}
\def\onehf            {{\textstyle \frac{1}{2} }}
\def\oneth            {{\textstyle \frac{1}{3} }}
\def\twoth            {{\textstyle \frac{2}{3} }}
\def\onefo            {{\textstyle \frac{1}{4} }}
\def\forth            {{\textstyle \frac{4}{3} }}

\newc{\matth}{\mathsurround=0pt}
\def\ML{\ifmmode{{\mathaccent"7E M}_L}
             \else{${\mathaccent"7E M}_L$}\fi}
\def\MR{\ifmmode{{\mathaccent"7E M}_R}
             \else{${\mathaccent"7E M}_R$}\fi}
\newcommand{\s}{\\ \vspace*{-3mm} }

\def \ud { {1 \over 2} }
\def \ut { {1 \over 3} }
\def \td { {3 \over 2} }
\newc{\mr}{\mathrm}
\def\dh {\partial }
\def \cs { cross-section }
\def \css { cross-sections }
\def \cm { centre of mass }
\def \cms { centre of mass energy }
\def \cc { coupling constant }
\def \ccs {coupling constants }
\def \gc {gauge coupling }
\def \gcc {gauge coupling constant }
\def \gccs {gauge coupling constants }
\def \yc {Yukawa coupling }
\def \ycc {Yukawa coupling constant }
\def \pp {{parameter }}
\def \pps {{parameters }} 
\def \ps {parameter space }
\def \pss {parameter spaces }
\def \vv {vice versa }

\newc{\siminf}{\mbox{$_{\sim}$ {\small {\hspace{-1.em}{$<$}}}    }}
\newc{\simsup}{\mbox{$_{\sim}$ {\small {\hspace{-1.em}{$>$}}}    }}


\newc {\Zboson}{{\mathrm Z}^{0}}
\newc{\thetaw}{\theta_W}
\newc{\mbot}{{m_b}}
\newc{\mtop}{{m_t}}
\newc{\sm}{${\cal {SM}}$}
\newc{\as}{\alpha_s}
\newc{\aem}{\alpha_{em}}
\def \PI{{\pi^{\pm}}}
\newc{\ppbar}{\mbox{$p\ol{p}$}}
\newc{\bbbar}{\mbox{$b\ol{b}$}}
\newc{\ccbar}{\mbox{$c\ol{c}$}}
\newc{\ttbar}{\mbox{$t\ol{t}$}}
\newc{\eebar}{\mbox{$e\ol{e}$}}
\newc{\zzero}{\mbox{$Z^0$}}
\def \gamz{\Gam_Z}
\newc{\wplus}{\mbox{$W^+$}}
\newc{\wminus}{\mbox{$W^-$}}
\newc{\ellp}{\ell^+}
\newc{\ellm}{\ell^-}
\newc{\elp}{\mbox{$e^+$}}
\newc{\elm}{\mbox{$e^-$}}
\newc{\elpm}{\mbox{$e^{\pm}$}}
\newc{\qbar}     {\mbox{$\ol{q}$}}
\def \ewgroup{SU(2)_L \otimes U(1)_Y}
\def \smgroup{SU(3)_C \otimes SU(2)_L \otimes U(1)_Y}
\def \smcolorem{SU(3)_C \otimes U(1)_{em}}

\def \SSM  {Supersymmetric Standard Model}
\def \poincare{Poincare$\acute{e}$}
\def \superspace{\emph{superspace}}
\def \sfs{\emph{superfields}}
\def \superpot{\emph{superpotential}}
\def \csf{\emph{chiral superfield}}
\def \csfs{\emph{chiral superfields}}
\def \vsf{\emph{vector superfield }}
\def \vsfs{\emph{vector superfields}}
\newc{\Ebar}{{\bar E}}
\newc{\Dbar}{{\bar D}}
\newc{\Ubar}{{\bar U}}
\newc{\susy}{{{SUSY}}}
\newc{\msusy}{{{M_{SUSY}}}}

\def\photino{\ifmmode{\mathaccent"7E \gam}\else{$\mathaccent"7E \gam$}\fi}
\def\taugluino{\ifmmode{\tau_{\mathaccent"7E g}}
             \else{$\tau_{\mathaccent"7E g}$}\fi}
\def\mphotino{\ifmmode{m_{\mathaccent"7E \gam}}
             \else{$m_{\mathaccent"7E \gam}$}\fi}
\newc{\gl}   {\mbox{$\wt{g}$}}
\newc{\mgl}  {\mbox{$m_{\gl}$}}
\def \charginopm{{\wt\chi}^{\pm}}
\def \mcharginopm{m_{\charginopm}}
\def \mchpmmin {\mcharginopm^{min}}
\def \chonep {{\wt\chi_1^+}}
\def \chone {{\wt\chi_1}}
\def \ch2p {{\wt\chi_2^+}}
\def \chonem {{\wt\chi_1^-}}
\def \ch2m {{\wt\chi_2^-}}
\def \chplus {{\wt\chi^+}}
\def \chminus {{\wt\chi^-}}
\def \chonip{{\wt\chi_i}^{+}}
\def \chonim{{\wt\chi_i}^{-}}
\def \chonipm{{\wt\chi_i}^{\pm}}
\def \chonjp{{\wt\chi_j}^{+}}
\def \chonjm{{\wt\chi_j}^{-}}
\def \chonjpm{{\wt\chi_j}^{\pm}}
\def \chonepm{{\wt\chi_1}^{\pm}}
\def \chonemp{{\wt\chi_1}^{\mp}}
\def \mchonepm{m_{\chonepm}}
\def \mchonemp{m_{\chonemp}}
\def \chtwopm{{\wt\chi_2}^{\pm}}
\def \mchtwopm{m_{\chtwopm}}
\newc{\dmchi}{\Delta m_{\wt\chi}}


\def \vlsp{\emph{VLSP}}
\def \lspi{\wt\chi_i^0}
\def \mlspi{m_{\lspi}}
\def \lspj{\wt\chi_j^0}
\def \mlspj{m_{\lspj}}
\def \lspone{\wt\chi_1^0}
\def \mlspone{m_{\lspone}}
\def \lsptwo{\wt\chi_2^0}
\def \mlsptwo{m_{\lsptwo}}
\def \lspthree{\wt\chi_3^0}
\def \mlspthree{m_{\lspthree}}
\def \lspfour{\wt\chi_4^0}
\def \mlspfour{m_{\lspfour}}


\newc{\sele}{\wt{\mathrm e}}
\newc{\sell}{\wt{\ell}}
\def \msell{m_{\sell}}
\def \slepone{\wt\ell_1}
\def \mslepone{m_{\slepone}}
\def \smuone{\wt\mu_1}
\def \msmuone{m_{\smuone}}
\def \stauone{\wt\tau_1}
\def \mstauone{m_{\stauone}}
\def \snu{\wt{\nu}}
\def \snutau{\wt{\nu}_{\tau}}
\def \msnu{m_{\snu}}
\def \msnumu{m_{\snu_{\mu}}}
\def \barsnu{\wt{\bar{\nu}}}
\def \barsnul{\barsnu_{\ell}}
\def \snul{\snu_{\ell}}
\def \mbarsnu{m_{\barsnu}}
\newc{\snue}     {\mbox{$ \wt{\nu_e}$}}
\newc{\smu}{\wt{\mu}}
\newc{\stau}{\wt{\tau}}
\newc {\nuL} {\wt{\nu}_L}
\newc {\nuR} {\wt{\nu}_R}
\newc {\snub} {\bar{\wt{\nu}}}
\newc {\eL} {\wt{e}_L}
\newc {\eR} {\wt{e}_R}
\def \slepl{\wt{l}_L}
\def \mslepl{m_{\slepl}}
\def \slepr{\wt{l}_R}
\def \mslepr{m_{\slepr}}
\def \stau{\wt\tau}
\def \mstau{m_{\stau}}
\def \slepton{\wt\ell}
\def \mslepton{m_{\slepton}}
\def \mlhiggs{m_{h^0}}

\def \xr{X_{r}}

\def \sfer{\wt{f}}
\def \msfer{m_{\sfer}}
\def \sq{\wt{q}}
\def \msq{m_{\sq}}
\def \msquleft{m_{\tilde{u_L}}}
\def \msqurht{m_{\tilde{u_R}}}
\def \sql{\wt{q}_L}
\def \msql{m_{\sql}}
\def \sqr{\wt{q}_R}
\def \msqr{m_{\sqr}}
\newc{\msqot}  {\mbox{$m_(\sq_{1,2} )$}}
\newc{\sqbar}    {\mbox{$\bar{\wt{q}}$}}
\newc{\ssb}      {\mbox{$\squark\ol{\squark}$}}
\newc {\qL} {\wt{q}_L}
\newc {\qR} {\wt{q}_R}
\newc {\uL} {\wt{u}_L}
\newc {\uR} {\wt{u}_R}
\def \ul{\wt{u}_L}
\def \mul{m_{\ul}}
\newc {\dL} {\wt{d}_L}
\newc {\dR} {\wt{d}_R}
\newc {\cL} {\wt{c}_L}
\newc {\cR} {\wt{c}_R}
\newc {\sL} {\wt{s}_L}
\newc {\sR} {\wt{s}_R}
\newc {\tL} {\wt{t}_L}
\newc {\tR} {\wt{t}_R}
\newc {\stb} {\ol{\wt{t}}_1}
\newc {\sbot} {\wt{b}_1}
\newc {\msbot} {m_{\sbot}}
\newc {\sbotb} {\ol{\wt{b}}_1}
\newc {\bL} {\wt{b}_L}
\newc {\bR} {\wt{b}_R}
\def \mul{m_{\wt{u}_L}}
\def \mur{m_{\wt{u}_R}}
\def \mdl{m_{\wt{d}_L}}
\def \mdr{m_{\wt{d}_R}}
\def \mcl{m_{\wt{c}_L}}
\def \charml{\wt{c}_L}
\def \mcr{m_{\wt{c}_R}}
\newc{\csquark}  {\mbox{$\wt{c}$}}
\newc{\csquarkl} {\mbox{$\wt{c}_L$}}
\newc{\mcsl}     {\mbox{$m(\csquarkl)$}}
\def \msl{m_{\wt{s}_L}}
\def \msr{m_{\wt{s}_R}}
\def \mbl{m_{\wt{b}_L}}
\def \mbr{m_{\wt{b}_R}}
\def \mtl{m_{\wt{t}_L}}
\def \mtr{m_{\wt{t}_R}}
\def \st{\wt{t}}
\def \mst{m_{\st}}
\newc {\stopl}         {\wt{\mathrm{t}}_{\mathrm L}}
\newc {\stopr}         {\wt{\mathrm{t}}_{\mathrm R}}
\newc {\stoppair}      {\wt{\mathrm{t}}_{1}
\bar{\wt{\mathrm{t}}}_{1}}
\def \lstop{\wt{t}_{1}}
\def \lstopbar{\lstop^*}
\def \hstop{\wt{t}_{2}}
\def \hstopbar{\hstop^*}
\def \mlstop{m_{\lstop}}
\def \mhstop{m_{\hstop}}
\def \lstoppair{\lstop\lstop^*}
\def \hstoppair{\hstop\hstop^*}
\newc{\tsquark}  {\mbox{$\wt{t}$}}
\newc{\ttb}      {\mbox{$\tsquark\ol{\tsquark}$}}
\newc{\ttbone}   {\mbox{$\tsquark_1\ol{\tsquark}_1$}}
\def \tsq {top squark }
\def \tsqs {top squarks }
\def \tsql {ligtest top squark }
\def \tsqh {heaviest top squark }
\newc{\mix}{\theta_{\wt t}}
\newc{\cost}{\cos{\theta_{\wt t}}}
\newc{\sint}{\sin{\theta_{\wt t}}}
\newc{\costloop}{\cos{\theta_{\wt t_{loop}}}}
\def \lsbot{\wt{b}_{1}}
\def \lsbotbar{\lsbot^*}
\def \hsbot{\wt{b}_{2}}
\def \hsbotbar{\hsbot^*}
\def \mlsbot{m_{\lsbot}}
\def \mhsbot{m_{\hsbot}}
\def \lsbotpair{\lsbot\lsbot^*}
\def \hsbotpair{\hsbot\hsbot^*}
\newc{\mixsbot}{\theta_{\wt b}}

\def \mhone{m_{h_1}}
\def \hup{{H_u}}
\def \hdn{{H_d}}
\newc{\tb}{\tan\beta}
\newc{\cb}{\cot\beta}
\newc{\vev}[1]{{\left\langle #1\right\rangle}}

\def \abot{A_{b}}
\def \atop{A_{t}}
\def \atau{A_{\tau}}
\newc{\mhalf}{m_{1/2}}
\newc{\mzero} {\mbox{$m_0$}}
\newc{\azero} {\mbox{$A_0$}}

\newc{\lb}{\lam}
\newc{\lbp}{\lam^{\prime}}
\newc{\lbpp}{\lam^{\prime\prime}}
\newc{\rpv}{{\not \!\! R_p}}
\newc{\rpvm}{{\not  R_p}}
\newc{\rp}{R_{p}}
\newc{\rpmssm}{{RPC MSSM}}
\newc{\rpvmssm}{{RPV MSSM}}


\newc{\sbyb}{S/$\sqrt B$}
\newc{\pelp}{\mbox{$e^+$}}
\newc{\pelm}{\mbox{$e^-$}}
\newc{\pelpm}{\mbox{$e^{\pm}$}}
\newc{\epem}{\mbox{$e^+e^-$}}
\newc{\lplm}{\mbox{$\ell^+\ell^-$}}
\def \branch{\emph{BR}}
\def \branche{\branch(\lstop\ra be^{+}\nu_e \lspone)\ti \branch(\lstop^{*}\ra \bar{b}q\bar{q^{\prime}}\lspone)}
\def \branchmu{\branch(\lstop\ra b\mu^{+}\nu_{\mu} \lspone)\ti \branch(\lstop^{*}\ra \bar{b}q\bar{q^{\prime}}\lspone)}
\def\Ecm{\ifmmode{E_{\mathrm{cm}}}\else{$E_{\mathrm{cm}}$}\fi}
\newc{\rts}{\sqrt{s}}
\newc{\rtshat}{\sqrt{\hat s}}
\newc{\gev}{\,GeV}
\newc{\mev}{~{\rm MeV}}
\newc{\tev}  {\mbox{$\;{\rm TeV}$}}
\newc{\gevc} {\mbox{$\;{\rm GeV}/c$}}
\newc{\gevcc}{\mbox{$\;{\rm GeV}/c^2$}}
\newc{\intlum}{\mbox{${ \int {\cal L} \; dt}$}}
\newc{\call}{{\cal L}}
\def \met  {\mbox{${E\!\!\!\!/_T}$}}
\def \cpv  {\mbox{${CP\!\!\!\!/}$}}
\newc{\ptmiss}{/ \hskip-7pt p_T}
\def \eslash{\not \! E}
\def \etslash{\not \! E_T }
\def \ptslash{\not \! p_T }
\newc{\PT}{\mbox{$p_T$}}
\newc{\ET}{\mbox{$E_T$}}
\newc{\dedx}{\mbox{${\rm d}E/{\rm d}x$}}
\newc{\ifb}{\mbox{${\rm fb}^{-1}$}}
\newc{\ipb}{\mbox{${\rm pb}^{-1}$}}
\newc{\pb}{~{\rm pb}}
\newc{\fb}{~{\rm fb}}
\newc{\ycut}{y_{\mathrm{cut}}}
\newc{\chis}{\mbox{$\chi^{2}$}}
\def \hadron{\emph{hadron}}
\def \nlc{\emph{NLC }}
\def \lhc{\emph{LHC }}
\def \cdf{\emph{CDF }}
\def\dzero{\emptyset}
\def \tevatron{\emph{Tevatron }}
\def \lep{\emph{LEP }}
\def \jets{\emph{jets }}
\def \jet(s){\emph{jet(s) }}

\def\Crs{stroke [] 0 setdash exch hpt sub exch vpt add hpt2 vpt2 neg V currentpoint stroke 
hpt2 neg 0 R hpt2 vpt2 V stroke}
\def\loopdk{\lstop \ra c \lspone}
\def\brloopdk{\branch(\loopdk)}
\def\fourdk{\lstop \ra b \lspone  f \bar f'}
\def\brfourdk{\branch(\fourdk)}
\def\fourdklep{\lstop \ra b \lspone  \ell \nu_{\ell}}
\def\fourdkhad{\lstop \ra b \lspone  q \bar q'}
\def\brfourdklep{\branch(\fourdklep)}
\def\brfourdkhad{\branch(\fourdkhad)}
\def\twodk{\lstop \ra b \chonep}
\def\brtwodk{\branch(\twodk)}
\def\threedkslep{\lstop \ra b \wt{\ell} \nu_{\ell}}
\def\brthreedkslep{\branch(\threedkslep)}
\def\threedksnu{\lstop \ra b \wt{\nu_{\ell}} \ell }
\def\brthreedksnu{\branch(\threedksnu) }
\def\threedklsp{\lstop \ra b W \lspone }
\def\brthreedklsp{\\branch(\threedklsp) }
\def\topdk{t \ra \lstop \lspone}
\def\rpvdk{\lstop \ra e^+ d}
\def\brrpvdk{\branch(\rpvdk)}
\def\fonec{f_{11c}} 
\newc{\mpl}{M_{\rm Pl}}
\newc{\mgut}{M_{GUT}}
\newc{\mw}{M_{W}}
\newc{\mweak}{M_{weak}}
\newc{\mz}{M_{Z}}

\newc{\OPALColl}   {OPAL Collaboration}
\newc{\ALEPHColl}  {ALEPH Collaboration}
\newc{\DELPHIColl} {DELPHI Collaboration}
\newc{\XLColl}     {L3 Collaboration}
\newc{\JADEColl}   {JADE Collaboration}
\newc{\CDFColl}    {CDF Collaboration}
\newc{\DXColl}     {D0 Collaboration}
\newc{\HXColl}     {H1 Collaboration}
\newc{\ZEUSColl}   {ZEUS Collaboration}
\newc{\LEPColl}    {LEP Collaboration}
\newc{\ATLASColl}  {ATLAS Collaboration}
\newc{\CMSColl}    {CMS Collaboration}
\newc{\UAColl}    {UA Collaboration}
\newc{\KAMLANDColl}{KamLAND Collaboration}
\newc{\IMBColl}    {IMB Collaboration}
\newc{\KAMIOColl}  {Kamiokande Collaboration}
\newc{\SKAMIOColl} {Super-Kamiokande Collaboration}
\newc{\SUDANTColl} {Soudan-2 Collaboration}
\newc{\MACROColl}  {MACRO Collaboration}
\newc{\GALLEXColl} {GALLEX Collaboration}
\newc{\GNOColl}    {GNO Collaboration}
\newc{\SAGEColl}  {SAGE Collaboration}
\newc{\SNOColl}  {SNO Collaboration}
\newc{\CHOOZColl}  {CHOOZ Collaboration}
\newc{\PDGColl}  {Particle Data Group Collaboration}

\def\issue(#1,#2,#3){{\bf #1}, #2 (#3)}
\def\ASTR(#1,#2,#3){Astropart.\ Phys. \issue(#1,#2,#3)}
\def\AJ(#1,#2,#3){Astrophysical.\ Jour. \issue(#1,#2,#3)}
\def\AJS(#1,#2,#3){Astrophys.\ J.\ Suppl. \issue(#1,#2,#3)}
\def\APP(#1,#2,#3){Acta.\ Phys.\ Pol. \issue(#1,#2,#3)}
\def\JCAP(#1,#2,#3){Journal\ XX. \issue(#1,#2,#3)} 
\def\SC(#1,#2,#3){Science \issue(#1,#2,#3)}
\def\PRD(#1,#2,#3){Phys.\ Rev.\ D \issue(#1,#2,#3)}
\def\PR(#1,#2,#3){Phys.\ Rev.\ \issue(#1,#2,#3)} 
\def\PRC(#1,#2,#3){Phys.\ Rev.\ C \issue(#1,#2,#3)}
\def\NPB(#1,#2,#3){Nucl.\ Phys.\ B \issue(#1,#2,#3)}
\def\NPPS(#1,#2,#3){Nucl.\ Phys.\ Proc. \ Suppl \issue(#1,#2,#3)}
\def\NJP(#1,#2,#3){New.\ J.\ Phys. \issue(#1,#2,#3)}
\def\JP(#1,#2,#3){J.\ Phys.\issue(#1,#2,#3)}
\def\JPG(#1,#2,#3){J.\ Phys. G.\issue(#1,#2,#3)}
\def\PL(#1,#2,#3){Phys.\ Lett. \issue(#1,#2,#3)}
\def\PLB(#1,#2,#3){Phys.\ Lett.\ B  \issue(#1,#2,#3)}
\def\ZP(#1,#2,#3){Z.\ Phys. \issue(#1,#2,#3)}
\def\ZPC(#1,#2,#3){Z.\ Phys.\ C  \issue(#1,#2,#3)}
\def\PREP(#1,#2,#3){Phys.\ Rep. \issue(#1,#2,#3)}
\def\PRL(#1,#2,#3){Phys.\ Rev.\ Lett. \issue(#1,#2,#3)}
\def\MPL(#1,#2,#3){Mod.\ Phys.\ Lett. \issue(#1,#2,#3)}
\def\RMP(#1,#2,#3){Rev.\ Mod.\ Phys. \issue(#1,#2,#3)}
\def\SJNP(#1,#2,#3){Sov.\ J.\ Nucl.\ Phys. \issue(#1,#2,#3)}
\def\CPC(#1,#2,#3){Comp.\ Phys.\ Comm. \issue(#1,#2,#3)}
\def\IJMPA(#1,#2,#3){Int.\ J.\ Mod. \ Phys.\ A \issue(#1,#2,#3)}
\def\MPLA(#1,#2,#3){Mod.\ Phys.\ Lett.\ A \issue(#1,#2,#3)}
\def\PTP(#1,#2,#3){Prog.\ Theor.\ Phys. \issue(#1,#2,#3)}
\def\RMP(#1,#2,#3){Rev.\ Mod.\ Phys. \issue(#1,#2,#3)}
\def\NIMA(#1,#2,#3){Nucl.\ Instrum.\ Methods \ A \issue(#1,#2,#3)}
\def\JHEP(#1,#2,#3){J.\ High\ Energy\ Phys. \issue(#1,#2,#3)}
\def\EPJC(#1,#2,#3){Eur.\ Phys.\ J.\ C \issue(#1,#2,#3)}
\def\RPP (#1,#2,#3){Rept.\ Prog.\ Phys. \issue(#1,#2,#3)}
\def\PPNP(#1,#2,#3){ Prog.\ Part.\ Nucl.\ Phys. \issue(#1,#2,#3)}
\newc{\PRDR}[3]{{Phys. Rev. D} {\bf #1}, Rapid  Communications, #2 (#3)}

\vspace*{\fill}
\vspace{-1.5in}
\begin{flushright}
{\tt IISER/HEP/04/11}
\end{flushright}
\begin{center}
{\Large \bf
SUSY signals with small and large trilinear couplings at the LHC 7 TeV runs
and neutralino dark matter }
  \vglue 0.4cm
  Nabanita Bhattacharyya\footnote{nabanita@iiserkol.ac.in},
  Arghya Choudhury\footnote{arghyac@iiserkol.ac.in} and
  Amitava Datta\footnote{adatta@iiserkol.ac.in}
      \vglue 0.1cm
          {\it 
	  Indian Institute of Science Education and Research, Kolkata, \\
          Mohanpur Campus, PO: BCKV Campus Main Office,\\
          Mohanpur - 741252, Nadia, West Bengal.\\
	  }
	  \end{center}
	  \vspace{.1cm}

\begin{abstract}

We propose the signal $1b+1l+N_j+\etslash$ along with appropriate selection 
criteria for the LHC 7 TeV run, where the number of jets $(N_j)$ is $\ge2$ or 4.
These signals can complement the canonical $Jets$ + $\etslash$ signature since 
they are sensitive to the trilinear soft breaking parameter $(A_0)$ and low values
of the parameter tan$\beta$ in the minimal supergravity (mSUGRA) model.
A large region of this mSUGRA parameter space within the reach of the ongoing 
experiments at the LHC is disfavoured by the bound on the lightest Higgs boson mass
($m_h \ge 114.4 $ GeV) unless $A_0$ has moderate to large negative values. 
Interestingly part of 
this parameter space with $A_0 \ne 0$ is also consistent with 
the observed dark matter relic density. 
A natural consequence of large $A_0$ is the existence of a light 
top squark ($\lstop$). The proposed signals primarily stem from direct 
$\lstop \lstop^*$ production  and/or $\tilde g \ra \tilde t_{1} t $, if all 
squark-gluino events are considered. A thorough analysis of the signals and 
the corresponding backgrounds are presented using the event generator Pythia.
We finally compare the signal size for $A_0 = 0$ and $A_0 \ne 0$.

\vspace{2.5 cm}

\end{abstract}


\section{Introduction}

The attention of the high energy physics community has been focussed on
the prospects of new physics search at the CERN Large Hadron Collider
(LHC) \cite{review}. Supersymmetry (SUSY) \cite{susy} is perhaps the 
most well-motivated extension of the standard model (SM) which will 
be extensively scrutinized under the LHC microscope.

It is gratifying to note that the proton - proton collisions with stable 
beams became operational in 2010 at an energy ($\sqrt{s}$ = 7 TeV) never 
attained by any accelerator before. Moreover, the performance of the 
machine in the luminosity front during 2010 surpassed all expectations. 
As a result it now expected that the operations will continue till the 
end of 2012 at $\sqrt{s}$ = 7 TeV and the total luminosity collected is 
expected to be even of the order of 5 $\ifb$. This is indeed an exciting news 
for the new particle search programme.

In some recent analyses \cite{7tev,naba2} the prospect of SUSY search at 
$\sqrt s = 7$ TeV have been studied. It has already been shown by both 
ATLAS \cite{atlas} and CMS \cite{cms} collaborations that even with 
$\lum =$ 1 $\ifb$ the minimal supergravity (mSUGRA) model \cite{msugra} 
can be probed in the $Jets$ + missing energy ($\etslash$) channel much 
beyond the reach of similar searches at the Tevatron \cite{teva}. 
Nevertheless the reach at 7 TeV will be restricted to relatively low 
values of $m_0$ and $m_{1/2}$.

Both the above analyses as well as many existing parameter space scans 
were carried out  for a specific choice of
the trilinear soft breaking term: $A_0 = 
0$.  However, it must be
admitted that there is no compelling reason for the  above choice.
The mSUGRA parameter space with $A_0 \ne 0 $ should be probed with 
equal emphasis after taking into account the constraints from the 
charge color breaking (CCB) minima \cite{CCB}. Unfortunately
$Jets$ + $\etslash$ signal - the main channel for SUSY search is rather 
insensitive to $A_0 $.

In this paper we wish to focus on the characteristic signatures of the 
mSUGRA model sensitive to non vanishing trilinear couplings ($A_0 \neq 
0$). This will complement the $Jets$ + $\etslash$ signature and provide 
handles on hitherto unexplored regions of the parameter space. There are 
several motivations for extending the analyses to this case. 

\begin{itemize} \item It is well known \cite{cmstdr} that a significant 
fraction of the low $m_0 - m_{1/2}$ region of the mSUGRA parameter space 
sensitive to the LHC 7 TeV run with relatively low luminosities is 
disfavoured by the bound $m_h > 114.4$ GeV, where $m_h$ is the mass of 
the lighter scalar Higgs boson, obtained from LEP \cite{hlim}. This is 
especially so if the mSUGRA parameter tan$\beta$ (to be defined below) 
is low. In contrast the small $m_0 - m_{1/2}$ region indeed opens up 
even for low tan$\beta$ for moderate to large negative values of $A_0$ 
\cite{debottam} which yields larger radiative corrections to enhance the 
predicted $m_h$. This point will be taken up in further details in the 
next section.\footnote{The whole parameter space accessible to the 7 TeV 
run can be made consistent with the $m_h$ bound for $A_0$ = 0, if the 
mSUGRA model is extended, e.g., to contain four sequential generations 
of fermions and sfermions. Due to additional radiative corrections to 
$m_h$ from the fourth generation the theoretical prediction 
significantly increases \cite{dawson}.} 
\item From the point of view of 
SUSY dark matter (DM) \cite{DM}, the zone of the parameter space thus 
revived is interesting \cite{debottam} since in this zone neutralino 
annihilation (bulk annihilation) and/or neutralino-stau co-annihilation 
can explain the observed DM relic density of the universe as given by 
WMAP data \cite{wmap}. In contrast for 
$A_0 = 0$, the parameter space compatible with the $m_h$ bound and DM 
data is restricted for $m_{1/2} \gsim 500 $. 

\item For small $m_0$ and $ m_{1/2}$ the signals involving 
isolated electrons and muons are sensitive to low values of tan$\beta$. 
For larger values, the mixing in the stau sector increases leading to a 
mass eigenstate ($\stau_1$) much lighter than the other sleptons. 
is especially true in the low $m_0 - m_{1/2}$ region of the parameter 
space. As a result the electroweak gauginos decay dominantly into 
$\stau_1$ and final states involving isolated e and/or $\mu$ are 
suppressed. It may be recalled that the D$\emptyset$ collaboration has 
obtained the best limit on the chargino-neutralino sector via the clean 
trilepton channel \cite{D0} for low tan$\beta$ (= 3). However, they have 
also taken $A_0 = 0$. As a results the entire parameter space sensitive 
to their search is strongly disfavoured by the $m_h$ bound from LEP.  
\end{itemize}

The phenomenology of models with  non-zero $A_0$ was discussed in
details in \cite{debottam} and subsequently in \cite{naba2,naba1}
for both LHC 7 TeV and 14 TeV runs.The 
emphasis of our work is  on SUSY signals at LHC 7 TeV, which are 
sensitive to $A_0$ and low values of tan$\beta$. We shall also compare 
and contrast them with the corresponding signals for $A_0 = 0$ .
A natural consequence of large negative $A_0$ is the existence of a  
lighter top squark mass eigenstate ($\lstop$). 
It will therefore be important to look for the footprints 
of this squark in the LHC 7 TeV data in addition to the canonical 
$Jets + \etslash$ channel. The  lighter top squark ($\lstop$)
can be copiously produced in two ways:
\begin{itemize}
\item Through direct $\lstop \lstop^*$ pair production. This signal
could be important, e.g. , for reconstructing the top squarks mass using
standard techniques.
\item Through all possible squark-gluino pair  ($\tilde g 
\tilde g$, $\tilde q \tilde g$ and  $\tilde q \tilde q$)  production
followed by the decay $\tilde g \ra \lstop^*$$(\lstop)t(\bar t)$.    
\end{itemize}

The above gluino decay has a large branching ratio (BR) over a large 
parameter space even if the $\tilde g $ is heavier than all squarks, 
since $\lstop$ could be much lighter than the other squarks as a consequence 
of  large $A_0$. For small tan$\beta$ the bottom squarks will be 
almost as heavy as the other squarks.  
Another  source of $\lstop$ is the decay
$\tilde b_1 \ra \lstop W$ where the initial $\tilde b_1$'s are obtained in the
final states either from gluino decays ($\tilde g \ra b \tilde b_1$)
or direct pair production. 

We shall look into the following issues: 
\begin{itemize}
\item The search prospect
of the signal from direct $\lstop \lstop^*$ production if all other
strongly interacting sparticles are beyond the reach of the 7 TeV run for low 
integrated luminosities. We shall propose a signal sensitive to
moderate to large values of $A_0$ and low tan$\beta$. 

\item We next consider all squark-gluino pair production in 
mSUGRA including
 $\lstop \lstop^*$ production. If the total signal stands over the 
background, we shall look into the possibility of separating the events 
due to pure 
stop pair production from the squark-gluino events. by introducing 
additional cuts.  If the purity of the stop sample thus obtained, is 
reasonable, one can possibly use the standard techniques of sparticle mass 
measurement \cite{mt2} to study the properties of the stop.

\item Finally in order to exploit the large missing energy 
in a typical SUSY signal, we shall consider  typical squark-gluino
events subject to stronger cuts designed to eliminate the SM background.
However, for reasons discussed above, even this signal  will be rich in
events containing one or more $\lstop$. We shall also compare 
this signal with the corresponding one for $A_0 = 0$. However 
filtering out  $\lstop \lstop^*$ sample with a small contamination from
squark-gluino events may not be feasible in this case .
\end{itemize}

The plan of the paper is as follows. In Section 2 we motivate a 
promising $1l + 1b + N_j + $ $\etslash$ signature sensitive 
to the magnitude of $A_0$ and small tan$\beta$, where $l = e, \mu$ 
and $N_j$ stands for the number of jets, by introducing a set of 
benchmark points. In Section 3 
we present a thorough analysis of the signal for $N_j\ge$ 2 or 4  and the  
corresponding SM backgrounds by using the event 
generator Pythia (version 6.409) \cite{pythia} and other supporting 
computational tools.  
Our conclusions and future outlooks are summarized in Section 4.

\section {The Benchmark Scenarios}

As discussed in the last section a possible way of identifying the 
regions of the mSUGRA parameter space with 
moderate to large negative values of the 
$A_0$ parameter is to look for the remnants of the lighter top squark
in the LHC  data. This of course depends on the  decay mode of $\lstop$.
  
If $\lstop$ happens to be the next to the lightest superparticle (NLSP) 
its dominant decay modes will be $\lstop \ra c \lspone$ \cite{hikasa}
and $\lstop \ra b \lspone f \bar f^{\prime}$ \cite{djouadi}.
The latter is particularly important at low tan$\beta$. From 
the simulations at Tevatron energies it seems that the  final states 
do not contain very hard particles while the missing energy \cite{manas}
is relatively low . Consequently,the signals will be rather 
difficult to detect. The competition between the two modes 
further adds to this 
difficulty. We have checked that this is more or less true even at the LHC 
7 TeV run. 

Moreover, $\lstop$ NLSP is not very common in 
mSUGRA. We shall not pursue this difficult search channel 
in this paper. A new channel for  
stop NLSP search  has been  proposed recently \cite{drees}.      

If $\lstop$ is heavier than the lighter chargino ($\chone^{\pm}$) ,
the mode $\lstop \ra b \chonep$ 
and its charge conjugate process will be the main decay mode 
for the 7 TeV run. However, this mode may compete with the decay 
$\lstop \ra t \lspone$ for some regions of the parameter space 
corresponding to relatively heavy $\lstop$. We shall
include both the competing modes in our analysis.

 If there is no sfermion lighter than the chargino (i.e., in the 
chargino NLSP scenario), the latter decays into final states $l \nu 
\lspone$ and $q q^{\prime} \lspone$, where the fermion pair may come 
either from a real or virtual $W$ (two body and three body decays). 
Sometimes the 
BRs of the leptonic three body decays of the chargino may be enhanced 
relative to the corresponding W decay  BRs due to additional 
contributions from virtual slepton exchanges. This happens if  
there are 
relatively light sleptons in the spectrum. For small tan$\beta$ 
the BR of these  decays involving the $\tau$'s will be roughly the
same as that for $e$ and $\mu$ channels. 
Some of the benchmark scenarios illustrate these points.

There are two interesting signatures from the decay cascades sketched 
above which are initiated by the $\lstop$ decays: $b~l~j\etslash$ and 
$l^+l^-j\etslash$ , where j stands for any number of jets. But the 
dilepton channel will be twice suppressed by the small leptonic BR. We 
have, therefore, looked for the $1l$ ($e$ or $\mu$) +$ 1b $jet(tagged) + 
$N_{j}$ (including untagged b jets) $\ge 2 + \etslash$ signal.

The $\chone^{\pm}$ may also decay directly 
into the two body mode  $\stau_1 + \nu_{\tau}$  
with large BR. This, however, is  more probable for moderate and 
large values of 
tan$\beta$ and relatively low $m_0$ and $m_{1/2}$, 
where $\stau_1$ can be naturally light. For establishing this signal 
tagging of the $\tau$ jets is 
required. In this paper we restrict ourselves to low tan$\beta$ and the 
above signal has not been considered. But some 
benchmark points indicate that  this decay may occur with large BR 
even for low tan$\beta$. We have checked that the proposed signal 
is viable inspite of this. 
The importance of the signal involving $\tau$ jets  
has been highlighted elsewhere in the context of the 14 TeV run 
\cite{debottam,naba1}. A similar analysis for the ongoing run is in 
progress.

The minimal supersymmetric extension of the standard model (MSSM) has 
too many free parameters due to unknown soft supersymmetry breaking 
terms. In this paper most of the analyses have been done within the simplest 
gravity mediated SUSY breaking model - mSUGRA 
\cite{msugra} model \footnote{The sub-section 3.1 is an exception.} -
which has only five free parameters including soft SUSY breaking terms. 
These are $m_0$ (the common scalar mass),
$m_{1/2}$ (the common gaugino mass), $A_0$ (the common trilinear coupling
parameter), all given at the gauge coupling unification scale 
($M_G \sim 2 \times 10^{16}$ GeV); the ratio of Higgs vacuum expectation 
values at the electroweak scale namely tan$\beta$ and the sign of $\mu$.
The magnitude of $\mu$ is determined by the radiative electroweak 
symmetry breaking (REWSB) 
condition \cite{rewsb}. The low energy sparticle spectra and couplings at the
electroweak scale are generated by renormalization group evolutions (RGE)
of the soft breaking masses and the coupling parameters.

Keeping the above features of the signal in mind we have chosen 
several benchmark point
(BP)s in  the mSUGRA parameter space. We have considered low, moderate 
and large values of the parameters $m_0$ and $m_{1/2}$ within the reach of
LHC 7 TeV run with 1 $\ifb$ of data as indicated by references 
\cite{atlas,cms}. However, we have considered  large to moderate  
negative values of the trilinear coupling $A_0$, which  yield a 
relatively light 
top squark and consistency with the $m_h$ bound from LEP. We have
fixed tan$\beta$ at 5 and have taken the sign of $\mu$ to be positive. 
In the following all parameters having dimensions of mass are in GeV unless 
stated otherwise explicitly.
For all the benchmark points  
$\lspone$ is a bino whereas $\chonep$ and $\lsptwo$ are winos 
with masses approximately equal to $M_2$. The pole masses of the top and 
the bottom quarks are taken to be 175 \footnote{We have made this choice to be consistency with 
\cite{debottam} as we have taken the WMAP allowed regions from this 
analysis. Admittedly this choice is slightly higher than the central
value given by PDG(2010): $ m_t = 172 \pm 0.9 \pm 1.3$. However, our
choice is well within the 2$\sigma$ limit of the central value which
is acceptable. We have checked that there is no major modification
in our numerical results due to this.} and 4.25 respectively.
The mSUGRA parameters for the  benchmark scenarios have been listed in 
Table 1. The sparticle spectra for the above scenarios are contained in 
Table 2 and 
Table 3. They have been generated by SUSPECT (version.2.3) \cite{suspect}.
\begin{table}[!htb]
\begin{center}\

\begin{tabular}{|c|c|c|c|}
\hline
             Benchmark   &   \multicolumn{3}{c|}{Parameters}\\
\cline{2-4}
        Points           & $m_0 $  &  $m_{1/2}$  & -$A_0$\\
       \hline
        BP1              & 130     &   195       &   600 \\
        BP2              & 150     &   195       &   650 \\
        BP3              & 350     &   195       &   650 \\
        BP4              & 450     &   195       &   900 \\
        BP5              & 115     &   195       &   600 \\
        BP6              & 115     &   235       &   600 \\
        BP7              & 115     &   285       &   600 \\
        BP8              & 150     &   179       &   600 \\
        BP9              & 150     &   215       &   600 \\
        BP10             & 290     &   205       &   600 \\
        BP11             & 290     &   235       &   600 \\
        BP12             & 290     &   285       &   600 \\
        BP13             & 490     &   235       &   600 \\
        BP14             & 490     &   275       &   600 \\

       \hline

       \end{tabular}
       \end{center}
          \caption{The benchmark scenarios with different  
$m_0$, $m_{1/2}$ and $A_0$ for fixed tan$\beta = 5$ and $sign(\mu) > 0$. In this 
paper all parameters with dimension of mass are in GeV unless stated otherwise 
explicitly. For the rationale of these choices see the text.}
          \end{table}

\begin{table}[!ht]
\begin{center}

\begin{tabular}{|c|c|c|c|c|c|c|c|}
       \hline
        Squark/     &		&	&	&      	&	&	&	\\
   Slepton/Gluino/  &	BP1  	&  BP2	& BP3	& BP4  	& BP5	& BP6	& BP7	\\
   Gaugino masses   &     	& 	&     	&    	&	&	&	\\

        \hline
$\wt g $	   &	485.8	& 487.3	& 500.2	& 507.4	& 484.9	& 574.2	& 685.4	\\
        \hline
$\wt q_L (\wt u_L)$&	461.5	& 467.6	& 557.8	& 621.7	& 457.5	& 537.2	& 636.7	\\
        \hline
$\wt q_R (\wt u_R)$&	449.8	& 456.1	& 549.3	& 614.5	& 445.6	& 521.9	& 617.3	\\
        \hline
$\wt t_1 $	   &	175.3	& 156.8	& 252.4	& 216.9	& 170.5	& 261.8	& 360.5	\\
        \hline
$\wt t_2 $	   &	494.9	& 496.1	& 545.8	& 573.8	& 492.7	& 521.9	& 641.8	\\
        \hline
$\wt b_1 $	   &	395.0	& 395.4	& 467.1	& 498.3	& 391.7	& 542.6	& 561.9	\\
        \hline
$\wt b_2 $	   &	450.1	& 456.2	& 548.6	& 612.9	& 445.9	& 521.8	& 615.8	\\
        \hline
$\wt l_L $	   &	190.7	& 204.8	& 375.0	& 468.8	& 180.9	& 201.0	& 227.9	\\
        \hline
$\wt \nu_{l_{L}} $   &	174.8	& 190.1	& 367.2	& 462.7	& 164.1	& 186.1	& 214.5	\\
        \hline
$\wt l_R $	   &	154.6	& 204.8	& 359.2	& 456.8	& 142.3	& 150.3	& 229.6	\\
        \hline
$\wt \tau_1 $	   &	146.5	& 163.7	& 354.7	& 451.4	& 133.5	& 142.6	& 154.9	\\
        \hline
$\wt \nu_{\tau_{L}}$ &	173.6	& 188.8	& 366.3	& 461.4	& 162.9	& 185.0	& 213.9	\\
        \hline
$\wt \tau_2 $	   &	193.8	& 207.8	& 376.7	& 470.3	& 184.2	& 203.5	& 229.6	\\
        \hline
$\chonepm $	   &	140.5	& 141.4	& 143.7	& 147.5	& 140.4	& 172.9	& 213.8	\\
        \hline
$\chtwopm $	   &	424.1	& 136.8	& 448.2	& 517.9	& 423.5	& 470.7	& 530.1	\\
        \hline
$\lspone $	   &	74.6	& 74.9	& 75.7	& 76.9	& 74.6	& 91.6	& 112.9	\\
        \hline
$\lsptwo $	   &	141.5	& 141.9	& 143.7	& 147.8	& 140.9	& 173.4	& 214.2	\\
        \hline
$ h $	  	   &	110.3	& 110.8	& 110.2	& 112.3	& 110.3	& 111.0	& 111.9	\\
        \hline
\end{tabular}
\end{center}
   \caption{The sparticle spectra for the benchmark points BP1 - BP7.}
\end{table}

\begin{table}[!ht]
\begin{center}

\begin{tabular}{|c|c|c|c|c|c|c|c|}
       \hline
        Squark/     &     	&	&	&	&	&     	&     	\\
   Slepton/Gluino/  & BP8	& BP9  	& BP10	& BP11	& BP12	& BP13	& BP14	\\
   Gaugino masses   &		&     	&	&	&	&     	&     	\\

        \hline
$\wt g $	   & 450.8	& 532.1	& 517.9	& 585.1	& 695.8	& 599.9	& 688.2	\\
        \hline
$\wt q_L (\wt u_L)$& 435.7	& 506.9	& 542.2	& 596.3	& 688.1	& 707.7	& 770.4	\\
        \hline
$\wt q_R (\wt u_R)$& 425.6	& 493.7	& 531.9	& 583.4	& 670.9	& 698.0	& 757.3	\\
        \hline
$\wt t_1 $	   & 139.8	& 228.1	& 257.1	& 310.6	& 397.2	& 391.8	& 448.6	\\
        \hline
$\wt t_2 $	   & 471.9	& 530.6	& 543.9	& 590.9	& 671.2	& 657.7	& 716.1	\\
        \hline
$\wt b_1 $	   & 369.0	& 437.5	& 461.5	& 513.4	& 600.9	& 599.2	& 660.6	\\
        \hline
$\wt b_2 $	   & 426.2	& 493.5	& 531.4	& 583.4	& 669.1	& 696.5	& 755.3	\\
        \hline
$\wt l_L $	   & 198.0	& 213.5	& 323.2	& 332.2	& 349.2	& 514.3	& 523.0	\\
        \hline
$\wt \nu_{l_{L}} $   & 182.7	& 199.5	& 314.2	& 323.5	& 340.9	& 508.7	& 517.6	\\
        \hline
$\wt l_R $	   & 169.3	& 174.9	& 302.2	& 305.2	& 310.9	& 498.4	& 501.2	\\
        \hline
$\wt \tau_1 $	   & 161.6 	& 167.9	& 297.6	& 300.8	& 306.9	& 495.0	& 497.9	\\
        \hline
$\wt \nu_{\tau_{L}}$ & 181.6	& 198.4	& 313.3	& 322.6	& 340.1	& 508.7	& 516.7	\\
        \hline
$\wt \tau_2 $	   & 201.3	& 216.0	& 324.9	& 333.6	& 350.2	& 514.9	& 523.5	\\
        \hline
$\chonepm $	   & 127.7	& 156.9	& 150.4	& 174.8	& 215.6	& 177.2	& 209.7	\\
        \hline
$\chtwopm $	   & 405.9	& 448.6	& 443.4	& 478.3	& 536.9	& 495.1	& 539.2	\\
        \hline
$\lspone $	   & 67.9 	& 83.2 	& 79.6	& 92.2	& 113.4	& 93.0	& 109.9	\\
        \hline
$\lsptwo $	   & 128.4	& 157.5	& 151.0	& 175.3	& 215.9	& 177.6	& 210.0	\\
        \hline
$ h $	  	   & 109.9	& 110.7	& 110.1	& 110.7	& 111.7	& 110.3	& 111.2	\\
        \hline
\end{tabular}
\end{center}
   \caption{ The sparticle spectra for the benchmark points BP8 - BP14.}
\end{table}
\begin{table}[!htb]
\begin{center}\
\begin{tabular}{|c|c|c|c|c|c|c|c|c|c|c|c|c|}
\hline
Channels                          	& BP1	& BP2	& BP3	& BP4	& BP5	& BP6	& BP7	\\
\hline
$\tilde g \ra \tilde t_{1} t    $     	& 48.1	& 55.5	& 88.9	& 99.5	& 45.6	& 39.6	& 35.8	\\
$\quad \ra \tilde b_{1} b    $     	& 23.4	& 23.8	& 11.0	& 0.46	& 22.7	& 20.7	& 18.7	\\

\hline

$\tilde q_{L} \ra \lsptwo q     $       & 33.0  & 33.0  & 24.7  & 17.3  & 33.0  & 32.9  & 32.8  \\
$\quad ~       \ra \chonepm q\prime $   & 66.4 & 66.4  & 49.6  & 34.7  & 66.3  & 66.1  & 65.9  \\
$\quad        \ra \tilde g q    $       & -     & -     & 24.9  & 47.4  & -     & -     & -     \\

\hline

$\tilde q_{R} \ra \lspone q     $       & 98.9  & 99.1  & 43.8  & 19.2  & 98.9  & 99.3  & 99.5  \\
$\quad        \ra \tilde g q    $       & -     & -     & 55.7  & 80.6  & -     & -     & -     \\

\hline

$\tilde t_{1} \ra \chonep b     $      	& 100.0	& 100.0	& 95.5	& 100.0	& 100.0	& 100.0	& 69.5	\\
$\quad        \ra \lspone t	$      	& --	& -	& 4.5	& -	& -	& -	& 30.5	\\

\hline

$\tilde b_{1} \ra \lspone b	$      	& 1.6	& 1.3	& 1.6	& 1.0	& 1.6	& 1.5	& 1.5	\\
$\quad        \ra \lsptwo b	$      	& 15.9	& 14.3	& 18.8	& 14.7	& 16.6	& 17.5	& 18.9	\\
$\quad        \ra \chonem t	$      	& 14.7	& 13.3	& 24.3	& 20.6	& 14.2	& 20.6	& 26.4	\\
$\quad        \ra \lstop W 	$      	& 67.8	& 71.0	& 55.3	& 63.6	& 68.6	& 60.3	& 53.2	\\

\hline\hline

$\chonepm \ra \lspone q q\prime $	& 44.4	& 52.2	& 67.4	& 67.5	& -	& -	& -	\\
$\quad    \ra \lspone l \nu_{l} $	& 33.0	& 30.5	& 21.7	& 21.6	& -	& -	& -	\\
$\quad    \ra \lspone \tau \nu_{\tau} $	& 16.5	& 17.3	& 10.8	& 10.8	& -	& -	&-	\\
$\quad    \ra\stau_{1}\nu_{\tau}$	& -	& -	& -	& -	& 100.0	& 98.6	& 74.6	\\
$\quad    \ra \lspone W        $	& -	& -	& -	& -	& -	& 1.9	& 25.4	\\

\hline

$\lsptwo \ra \lspone q \bar q  $	& -	& -	& 49.5	& 69.6	&  -	&-	&-	\\
$\quad   \ra \lspone b \bar b  $	& -	& -	& 23.5	& 24.2	& -	&-	&-	\\
$\quad   \ra \lspone \nu \bar\nu    $	& 46.0	& 47.6	& 4.5	& 1.4	& -	&-	&-	\\
$\quad   \ra \lspone \tau^+  \tau^- $	& 28.4	& 19.2	& 1.7	& 1.4	& -	&-	&-	\\
$\quad   \ra \stau_{1}^{\pm} \tau ^{\mp}$	& -	& -	& -	& -	& 100.0	& 89.4	& 85.4	\\
$\quad 	 \ra \lspone \l^+  \l^-   $	& 17.4	& 18.9	&2.8	&2.4	& -	& -	& -	\\
\hline

\end{tabular}
\end{center}
\caption{The BRs ($\%$) of the dominant decay modes of the gluino,
squarks and the electroweak gauginos for the benchmark points BP1 - BP7. Here $l$ = e and $\mu$.}
\end{table}

\begin{table}[!htb]
\begin{center}\
\begin{tabular}{|c|c|c|c|c|c|c|c|}
\hline
Channels                          	& BP8	& BP9	& BP10	& BP11 	& BP12	& BP13	& BP14	\\
\hline
$\tilde g \ra \tilde t_{1} t    $     	& 56.8	& 47.4	& 77.4	& 74.2	& 63.6	& 100.0	& 93.6	\\
$\quad    \ra \tilde b_{1} b    $      	& 25.4	& 23.2	& 22.6	& 25.6	& 24.7	& -	& 6.4	\\

\hline

$\tilde q_{L} \ra \lsptwo  q	   $   	& 33.0	& 32.9	& 30.7	& 32.4	& 32.8	& 19.8	& 24.3	\\
$\quad        \ra \chonepm q\prime $    & 66.4	& 66.2	& 61.7	& 65.1	&  65.7	& 39.8	& 48.9	\\
$\quad 	      \ra \tilde   g  q	   $   	& -	& -	& 6.5	& 1.3	& -	& 39.3	& 25.4	\\

\hline

$\tilde q_{R} \ra \lspone q	$    	& 98.7	& 99.2	& 88.4	& 99.3	& 99.5	& 26.2	& 44.3	\\
$\quad 	      \ra \tilde  g  q	$    	& -	& -	& 10.7	& -	& -	& 73.5	& 55.4	\\

\hline

$\tilde t_{1} \ra \chonep b     $      	& 100.0	& 100.0	& 94.4	& 77.1	& 62.9	& 58.4	& 53.2	\\
$\quad 	      \ra \lspone t	$      	& - 	& -	& 5.6	& 22.9	& 32.2	& 30.4	& 33.7	\\

\hline

$\tilde b_{1} 	\ra \lspone b	$      	& 1.6	& 1.6	& 1.6	& 1.6	& 1.4	& 1.6	& 1.4	\\
$\quad		\ra \lsptwo b	$      	& 15.3	& 17.1	& 19.1	& 19.6	& 67.1	& 22.9	& 22.2	\\
$\quad 		\ra \chonem t	$      	& 12.6	& 18.8	& 23.8	& 26.5	& 30.2	& 36.3	& 36.2	\\
$\quad 		\ra \lstop  W 	$      	& 70.4	& 62.5	& 55.4	& 52.2	& 48.2	& 38.9	& 39.8	\\

\hline\hline

$\chonepm \ra \lspone q q\prime  $	& 53.5	& 54.7	& 67.1	& -	& -	& -	& -	\\
$\quad 	 \ra \lspone l \nu_{l}   $	& 30.1	& 28.6	& 21.8	&-	&-	&-	& -	\\
$\quad 	 \ra \lspone \tau \nu_{\tau} $	& 16.3	& 16.7	& 10.9	&-	&-	&-	& -	\\
$\quad 	 \ra\stau_{1}\nu_{\tau}  $	& -	& -	&-	&-	&-	&-	& -	\\
$\quad 	 \ra \lspone  W          $	& -	& -	&-	& 100.0	& 100.0	& 100.0	& 100.0	\\

\hline

$\lsptwo \ra \lspone q \bar q  $	& 12.4	& 11.3	& 65.0	& 65.8	&-	& 65.0	& -	\\
$\quad 	 \ra \lspone b \bar b  $	& 4.5	& 3.8	& 21.4	& 20.4	&-	& 19.7	& -	\\
$\quad 	 \ra \lspone \nu \bar\nu    $	& 48.3	& 43.2	& 6.2	& 5.1	&-	& 9.6	& -	\\
$\quad 	 \ra \lspone \tau^+  \tau^- $	& 16.2	& 22.7	& 3.4 	& 3.1	&-	& 1.8	& -	\\
$\quad 	 \ra \lspone \l^+ \l ^- $	& 18.4	& 18.8	&5.5	&5.1	&-	&3.5	& - 	\\
$\quad 	 \ra \lspone  Z   $		& -	& -	& -	& -	& 100.0	& -	& 100.0	\\

\hline

\end{tabular}
\end{center}
\caption{The same as in Table 4 but for points BP8-14.}
\end{table}

Most of our parameter sets are consistent with the bound 
$m_{\chone^{\pm}}$ 
$\ge$ 141   from 
Tevatron \cite{D0} although, strictly speaking, this bound is valid for 
$A_0 = 0 $ and tan$\beta$ = 3 and for a specific model which maximizes
the BRs of $\chonepm$ and $\lsptwo$ into $e$ and $\mu$ channels. 

It may also be noted that there are sizable differences in the radiative 
corrections to $m_h$ as computed by the different tools. For example 
ISAJET \cite{isajet} gives $m_h$ larger by about +2 for the same mSUGRA 
parameters. Moreover, there is a $\sim$ 3 correction to $m_h$ due to yet 
unknown higher order effects \cite{hinemeyer}. In view of these 
uncertainties, the values of $m_h$ as given in Tables 2-3 are compatible 
with the LEP bound.

The BRs of gluino, squarks and gauginos for the above BPs 
have been computed by SDECAY \cite{sdecay} and are presented in Table 4 
and Table 5.

As already noted the light stop signature can come both from direct stop 
pair production or via the decay of the gluino. In BP1 the direct stop 
pair production is sizable due to the presence of a very light stop 
($m_{\lstop} = 175.3$) and the bulk of the signal comes from them. In 
contrast for BP2 and BP8 the corresponding masses are  even lighter 
($m_{\lstop} =$ 156.8 and 139.8). Nevertheless 
the signal from direct $\lstop \lstop^*$ pair production is small. This 
is because of the small $\Delta m = m_{\lstop} - m_{\chonep}$, which 
reduces the $b$ jet tagging efficiency. In such cases the main signal 
comes from $\tilde g \ra \lstop t$. 

The enhanced leptonic BR of $\chonepm$, as discussed above, for 
BP1, BP2 and BP8 should be noted. We stress that if direct $\lstop 
\lstop^*$ pair production is the main source of the signal (see section 
3.1) 
this enhancement is necessary. In contrast if the signal stems from
squark-gluino production followed by $\tilde g \ra \lstop t$ (see 
section 3.2) then there 
are many sources of b-jets and leptons. As a result observable signals  
are possible even if $\chonepm$ decay into $\lspone~ W$ or even into
$\stau_1 \nu_{\tau}$ modes with large BRs.

In BP1, BP2, BP5-BP9 and BP12  $m_{\tilde g} > m_{\tilde q}$, where 
$\tilde q$ refers to squarks of both $L$ and $R$ types belonging
to the first two generations. Here $\tilde g$ decays into $\tilde qq $ pairs 
of all flavours although the decays into the third generation dominate. 
In BP3, BP4, BP10, BP11, BP13 and BP14, 
$m_{\tilde q} > m_{\tilde g}$. As a result the gluino decays exclusively 
into the third generation squark-quark pairs. Thus inspite of
somewhat heavier $\lstop$'s, the signal may come from 
squark-gluino events.

In BP5, BP6, BP7 and BP9 the BR($\tilde g \ra \lstop t$) 
is not very large as
the gluinos are heavier than all squarks. Moreover, 
the BR of the two body decay $\chonep \ra \stau_1 \nu_{\tau}$ is 
quite large: being 100$\%$,
$98.6\%$, $74.6 \%$ and 100$\%$ respectively.
Lepton in such cases comes from 
$t$ and, to a lesser extent, from $\tau$ decay.
The scenario BP5 with low $m_0 - m_{1/2}$ has a 
light $\lstop$ with mass $\approx 171$. But in BP6 and BP7 the $\lstop$ 
is significantly heavier.
In spite of these unfavourable features of the chosen parameters  we 
shall
show that the signal can be obtained provided the selection criteria
are carefully chosen.

In BP11, BP12, BP13 and BP14 the $\lstop$ is comparatively heavy with
mass between 300 - 450. Here the decay channel of $\lstop \ra t \lspone$ 
opens up and contribute to the $1l + 1b + \etslash$ signal. 

In summary, the observables like the production cross sections, the BRs 
of $\lstop$ and gluino decays, 
the leptonic BR of the lighter charginos decays are chosen such 
that they vary between favourable and unfavourable values. Yet as we 
shall see in section 3 the proposed signal is viable over a large 
parameter space.

We have  chosen tan$\beta = 5$ for reasons 
. But we will see in Section 3.2 that even after relaxing this
restriction, signals from squark-gluino events will still be 
observable at low integrated luminosities.

\section {The Signal and the Backgrounds}

In this analysis we have generated all squark-gluino events at $\sqrt{s} 
= 7$ TeV using Pythia \cite{pythia}. Initial and final state radiation,
decay, hadronization, fragmentation and jet formation are implemented
following the standard procedures in Pythia. The lowest order
squark-gluino production cross-sections have been computed by CalcHEP (version 2.5.6)
\cite{calchep}.

We have used the toy calorimeter simulation (PYCELL) provided in Pythia
with the following criteria:

\begin{itemize}
\item The calorimeter coverage is $\vert \eta \vert < 4.5$. 
The segmentation is given by $\Delta \eta \times \Delta \phi = 0.09 \times 0.09$ which
resembles a generic LHC detector.

\item A cone algorithm with $\Delta$ R$ = \sqrt {\Delta\eta^2 + \Delta\phi^2}= 0.5 $ 
has been used for jet finding.

\item $E^{jet}_{T,min} = 30$ and jets are ordered
in $E_T$.
\end{itemize}

\subsection{Signature of direct stop pair production}

\begin{figure}[tb]
\begin{center}
\includegraphics[width=\textwidth]{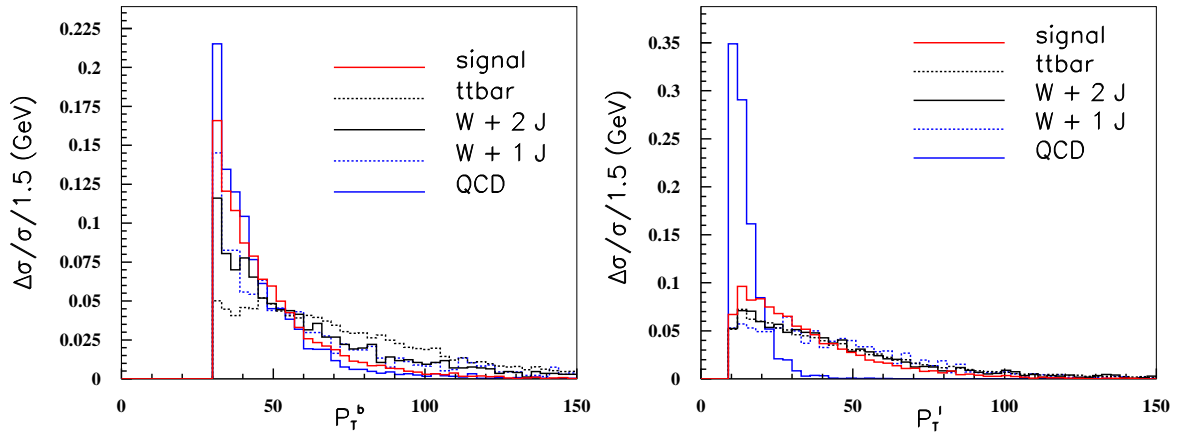}
\end{center}
\caption{The distributions (normalised to unity)
of $P_T$ of the {\it tagged} $b$-jet (left) and $P_T$ of
the isolated lepton (right) for $1b + 1l$ events (before the selection cuts) 
for the direct $\lstop \lstop^*$ signal and the dominant backgrounds. The SUSY spectrum 
is obtained from BP1.}
\label{fig1}
\end{figure}

\begin{figure}[tb]
\begin{center}
\includegraphics[width=\textwidth]{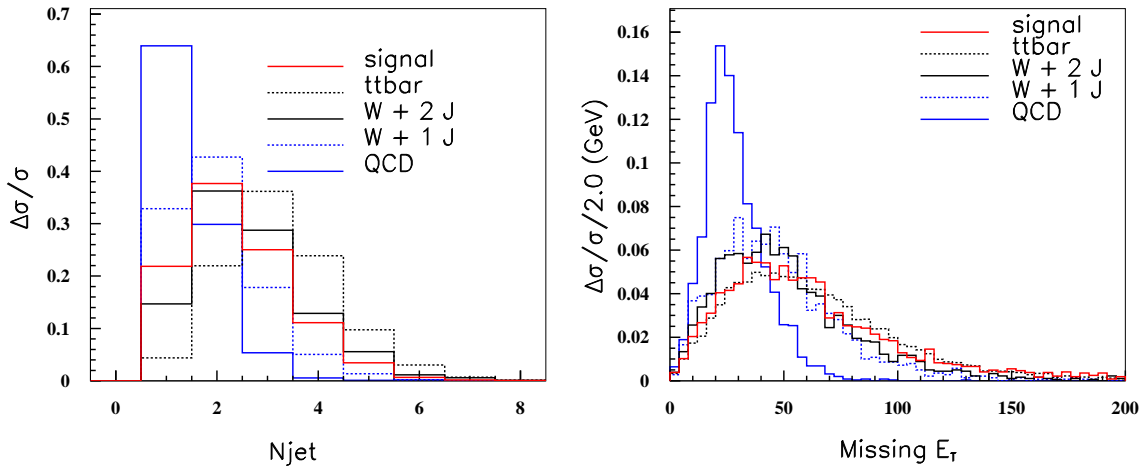}
\end{center}
\caption{The same as in Fig. 1 but for the distributions of 
number of jets (left) and $\etslash$ (right).}
\label{fig2}
\end{figure}

In this sub-section we have concentrated on the signature from $\lstop
\lstop^*$ pair production followed by cascade decays of both. This 
signature will be the main signature of SUSY if all strongly interacting 
sparticles except $\lstop$ are beyond the reach of LHC 7 TeV at low luminosities. 
Of course this scenario can not be realized in mSUGRA but is certainly possible in 
an unconstrained MSSM. It is, however, assumed that the masses and the BRs of the 
$\lstop$ and the sparticles in the electroweak sector are as in 
mSUGRA (See Tables 2 - 5). This ensures that bulk annihilation  
and/or stau-LSP co-annihilation produces the observed DM relic density \cite{debottam}.
We shall study in the next section the full consequences of mSUGRA when all 
sparticles are simultaneously produced . The point BP1 with the other
squarks and gluinos are assumed to be heavy
as discussed above, is a representative point for this analysis.

We have generated only $\lstop \lstop^*$ events 
using Pythia \cite{pythia}.  As already noted in Section 2 the BRs of the 
leptonic decays of the 
lighter chargino and the second lightest neutralino into leptonic 
channels containing the $e$ and $\mu$ in this case
are relatively high. 
This can naturally happen in any scenario with light sleptons.
This scenario motivates the signal $1l + 1b + N_j(\ge 2) +
\etslash$. We have required the following selection criteria:

\underline{Lepton selection:}

Leptons $(l=e,\mu)$ are selected with $P \mathrm{_T \ge 10}~  $
and $\vert\eta \vert < 2.4$. For lepton-jet isolation
we require $\Delta R(l,j) > 0.5$. For the sake of simplicity
the detection efficiency of $e$ and $\mu$ are assumed to be $ 100 \%$.

\underline{$b$- jet identification:}

We have tagged $b$-jets in our analysis by the following procedure.
A jet with $|\eta|< 2.5$ corresponding to the coverage of tracking detectors
matching with a $B$-hadron of decay length $> 0.9$ mm
has been marked $tagged$.
This criteria ensures that single $b$-jet tagging efficiency
(i.e., the ratio of tagged $b$-jets and the number of taggable $b$-jets)
$\epsilon_b \approx 0.5$ in $t \bar t$ events.
In Fig. 1 we have presented the $P_T$ distribution of the isolated
lepton (right) and the {\it tagged} $b$-jet (left)
for $\lstop \lstop^*$ events for BP1 scenario along with the dominant 
backgrounds.
 
 The following cuts, henceforth referred to as {\it Set 1},
 are implemented for background rejection :
\begin{itemize}

\item We have selected events with one isolated lepton ({\it cut 1}).

\item We have selected events with one {\it tagged} $b$ jet ({\it cut 2}).

\item Events with at least 2 jets are selected ({\it cut 3}). This is 
motivated by Fig. 2 (left). 

\item Events with missing transverse energy ($\etslash) \ge 75$   are 
selected ({\it cut 4}). It is to be noted that the signal has relatively 
low $\etslash$ (see Fig. 2 (right)).

\item We have also demanded events with $P_T$ {\it tagged} $b$ jet $\le 
80$   ({\it cut 5}). This is motivated by Fig. 1 (left)) and rejects
the background from $t \bar{t}$ events quite efficiently. 
\end{itemize}

We have considered the backgrounds from  $t \bar t$, QCD dijet production,
$W$ + $n$-$jets$ events, where $W$ decays into all channels.

We have generated $t \bar t$ events using Pythia and the leading order 
(LO) cross-section has been taken from CalcHEP which is 85.5 pb. We have generated
 QCD di-jet processes by Pythia in different $\hat p_T$ bins :
$25 \le \hat p_T \le 400$  , $400 \le \hat p_T \le 1000$   and 
$1000 \le \hat p_T \le 2000$  , where $\hat p_T$ is defined in
the rest frame of the parton parton collision.
The main contribution comes from the low $\hat p_T$ bin, 
which has a cross-section of $\sim 7.7 \times 10^7$ pb.
For the other bins the backgrounds are negligible.

For $W$ + $n$-$jets$ backgrounds we have generated events with $n=1$ and $2$
at the parton level using ALPGEN (version 2.13) \cite{alpgen}.
We have generated these events subject to the condition 
$P_T^j > 20$  , $\Delta R(j,j) \ge 0.3$ and
$\vert \eta \vert \le 4.5$.
These partonic events have been fed to Pythia for parton showering,
hadronization, fragmentation, decays etc.

All pair production  cross-sections (except for the QCD processes) are 
computed  in the leading order setting both the
renormalization and factorization scale equal,
$\mu_R = \mu_F = M$, where $M$ is the mass of the particle 
or sparticle produced and using CTEQ5L PDFs ~\cite{cteql}.
For example, for top and stop pair production the scale is set at
$m_t$ and $m_{\lstop}$ respectively. 
For QCD events the scale has been chosen to be
equal to $\sqrt {\hat s}$. For final states containing particles
having unequal masses the scale is set at the average mass of the
pair.

In Fig. 3 we delineate the regions on the $m_0 - A_0$ plane
which are accessible to the signal  from $\lstop\lstop^*$ pair 
production only corresponding to different integrated 
luminosities. The  parameters used to generate the stop and the 
electroweak sectors are $m_{1/2} = 195$, tan$\beta = 5$ 
and $sign(\mu) > 0$. 
As already mentioned other strongly interacting sparticles are assumed 
to be beyond the reach of LHC 7 TeV run. 

We have assumed that a signal is observable if  $S/\sqrt B \ge 5$, 
where $S$ ($B$) are the number of signal 
and background events respectively. In the pink region the signal from 
$\lstop \lstop^*$ events alone will stand over the background. For 
example, the number of
signal events after {\it Set 1} cuts, in the range $m_0 =$ 120 - 190, 
$m_{1/2} = 195$,
$A_0 =$ -560 to -640, tan$\beta = 5$ varies between 317 - 395 for 
$\lum =$ 1 $ \ifb$. 

The SM backgrounds for this set of cuts 
are given in Table 6. However, the signals presented in the first four 
columns of this table will be discussed in the next section.

\begin{figure}[!htb]
\begin{center}
\includegraphics[angle =270, width=1.0\textwidth]{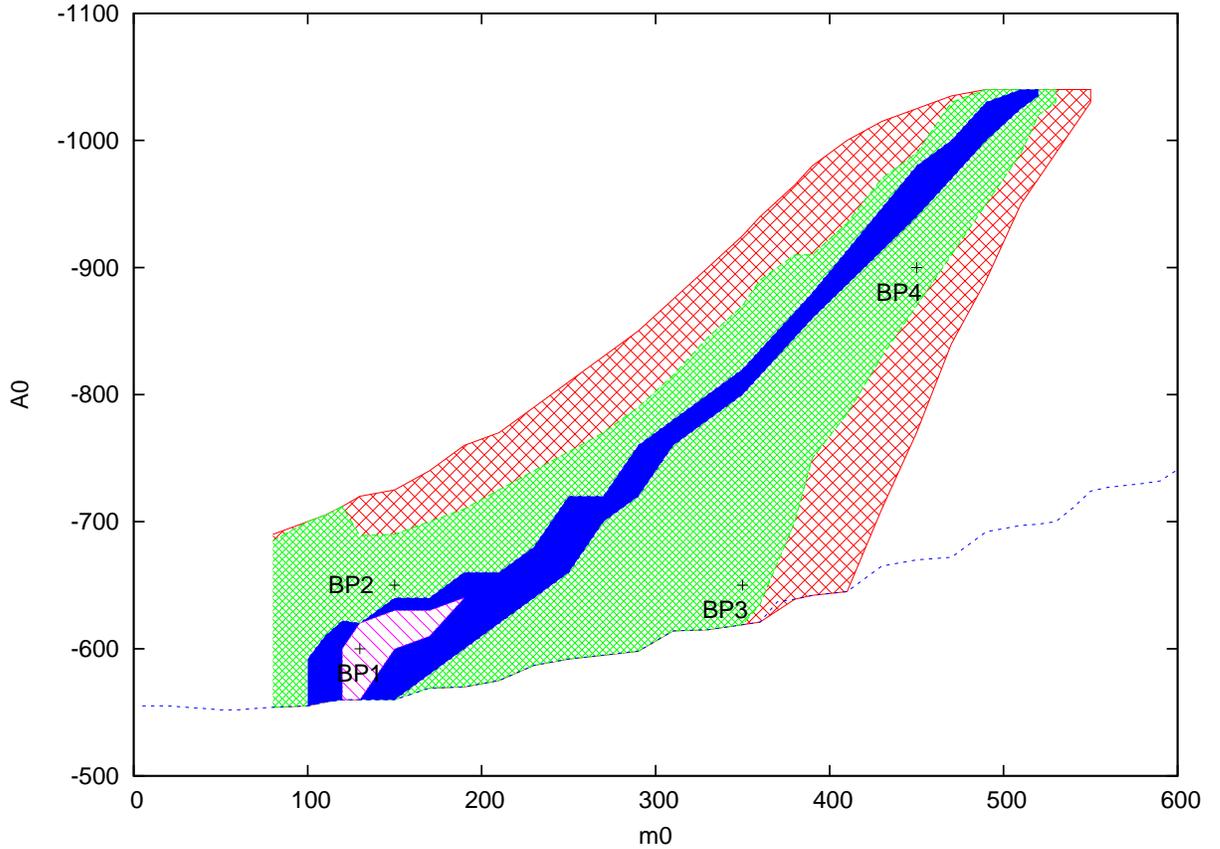}
\end{center}
\caption{Regions of the $m_0$ - $A_0$ plane with $m_{1/2} = 195$, tan$\beta =$ 5 and $sign(\mu) >$ 0
which can be probed by the $1l + 1b + N_j(\ge 2) + \etslash$ signal from direct 
$\lstop \lstop^*$ pair production only (Section 3.1) for $\lum =$ 1 (pink region)
and 2 (blue region) $\ifb$.The green (red) 
region is sensitive to the same signal 
from all squark-gluino events for $\lum =$ 1 (2) $\ifb$ (see Section 3.2).}
\end{figure}


Using the estimated backgrounds in Table 6, the significance
($S/\sqrt B$) for the above range varies between 5.0 to 6.2. This 
estimate is, however, conservative. If we compute the 
the next to leading order (NLO) cross-sections using PROSPINO
\cite{prospino}  and
CTEQ5M PDF we get a K- factor of 1.7. We do not know all the 
background cross-sections at the NLO level at 7 TeV. However, even if
our total background estimate is smaller by a factor of two (three), 
we would get a better (nearly equal) significance at NLO.

For  $\lum =$ 2 $\ifb$, regions corresponding to higher values of $m_0$
and $A_0$ become accessible to the 7 TeV run (see the blue region in Fig. 3). 
However, the stop mass reach does not change 
dramatically with increasing $\lum$. For $\lum =$ 1, 2 and 5 $\ifb$ the 
stop mass range that yields observable signal are 164 - 192 , 162 - 207
and 159 - 220 respectively. The lower edge of the above 
range is fixed by the magnitude of  $\Delta m = m_{\lstop} - 
m_{\chonep}$.

Since we require the $\chone^{\pm}$ to be lighter than the $\lstop$,
 with $m_{\chonepm} \gsim$ 141, the signal is visible for a narrow range 195   $\leq m_{1/2} \leq$ 215  .
The upper side of the range opens up for higher $\lum$.

The above results are not the most general ones since the mSUGRA type
correlations among  $m_{\lstop}$ and the masses in the electroweak 
sector 
have been assumed. In the most general MSSM,
the signal may be  visible over a larger parameter space. For example if 
the charginos dominantly decay leptonically via 2 body modes involving slepton-neutrino 
(or sneutrino-lepton) pairs with a nearly universal leptonic BR
and final state with one or more isolated leptons are required,  a much 
larger parameter space can be probed. Such scenarios may arise in 
the supergravity framework by varying the boundary conditions at the GUT 
scale in theoretically well motivated ways \cite{asesh}.

\subsection{Signals from Squark-Gluino events}

A much larger region of the parameter space can be scanned through all
squark-gluino events. 
We have generated these events at $E_{CM} =
7$ TeV using Pythia \cite{pythia}. The points BP1 - BP4 are the
representative points for this analysis. 

The efficiencies of the cuts belonging to {\it Set 1} for these points
are shown in Table 6 along with the SM background. The main source of 
$1l + 1b + N_j(\ge2) + \etslash$ signal is the decay
$\tilde g \ra \lstop t$.  In last row of Table 6 the significance of the signals 
for the selected points at $\lum =$ 1 $\ifb$ are shown. In Fig. 3 we
show the regions of the $m_0$ - $A_0$ plane with $m_{1/2}$ = 195,
tan$\beta$ = 5 and sign ($\mu$) $>$ 0 where the signal is visible  
for different integrated luminosities. For $\lum =$ 1 $\ifb$ (2 $\ifb$) 
the green (red) region yields observable signal. 
In Fig. 4 we have presented the same information
in the $m_0 - m_{1/2}$ plane for a fixed $A_0 (=-600  )$. The green (red) 
regions at the bottom of this figure corresponds to observable signals for 
$\lum =$ 1 $\ifb$ (2 $\ifb$).

\begin{table}[!htb]
\begin{center}\
\begin{tabular}{|c|c|c|c|c||c|c|c|c|}
\hline
             & \multicolumn{4}{c|}{Signal} & \multicolumn{4}{c|}{Background} \\
\hline
              & BP1   & BP2  & BP3 & BP 4 & $t \bar t$ &     QCD  & $W + 1j$    & $W + 2j$  \\
$\sigma$ (pb) & 26.7  & 36.9 & 8.0 & 8.4  &    85.5    & 7.7E+07  &  1.43E+04   &  5200     \\
\hline
\hline

{\it cut 1} & 9.2115 &12.9063  & 2.5783 & 2.85    &28.728     &  2.2E+05        &3066.8  &953.33  \\
\hline
{\it cut 2} &2.8417  &1.5424  & 1.0197  & 1.2017  & 14.0519   &  1.07E+04       & 8.2939 &  8.5925\\
\hline
{\it cut 3} & 2.4644 & 1.4464 & 0.9985  & 1.1261  & 13.4354   &   3847.4        & 5.571  & 7.3292 \\
\hline
{\it cut 4} &1.4524  & 1.1303 & 0.7480  & 0.6770  & 4.2382    &   8.6364        & 0.9473 & 1.5282 \\
\hline
{\it cut 5} &0.8908  & 0.5940 & 0.3460  & 0.3667  & 2.4615    &   0.0048        & 0.560   & 1.0078 \\
\hline
$S/\sqrt B$ &14.02   &9.35    &5.45   & 5.77 	  &   	      &                 &         &        \\

\hline
\end{tabular}
\end{center}
\caption{The total squark-gluino production cross-sections for different BPs
 and the SM backgrounds are at the top of the respective columns. 
The cross-sections after cuts of {\it Set 1} are presented step by step in rows 1-5.}
\end{table}

However, these events cannot be directly used for studying the 
properties of $\lstop$. We now apply additional cuts so that
the signal from direct stop pair production can be isolated from
the squark- gluino events. 

For the next part of the analysis we assume that the SM background can be accurately
estimated either from the data or by using improved higher order calculation
of the respective cross-sections and can be subtracted out.
In addition to the cuts belonging to {\it Set 1} we impose the following 
requirements on all squark-gluino events.   

\begin{itemize}
\item  $N_{central-jet} \le 4$, where central jets have $\vert \eta \vert \le$ 2.5 .
\item  The effective mass ($M_{eff}$)$\le 500$,
 where $M_{eff}= |\met| + \Sigma_{i}|P_T^{j_i}| + \Sigma_{i}|P_T^{l_i}| $
  ($l_i = e,\mu$ )
\end{itemize}

Both the above cuts eliminates bulk of the squark-gluino events while leaving 
the events from direct stop pair production relatively unaffected. 
These two cuts along with those belonging to cuts of {\it Set 1} define 
the {\it Set 2} of the cuts.
The results are summarized in Table 7 for different mSUGRA points taken 
from different regions of Fig. 3. The last two columns give 
number of all squark-gluino events after the cuts of {\it Set 1} and 
{\it Set 2} for
$\lum =$ 1 $\ifb$. The columns 3 and 4 give the corresponding 
number of events from stop pair events only. 
Comparing columns 4 and 6 we find that after {\it Set 2} cuts 
about 80 \%  of the remaining events are from $\lstop \lstop^*$  production.
It will be interesting to see whether  $\lstop$ mass can be reconstructed 
from such samples using the standard procedures \cite {mt2}. 
It is important to note that the purity of the $\lstop \lstop^*$ sample 
increases as the $\wt q$ and $\wt g$ masses increase.

\begin{figure}[!htb]
\begin{center}
\includegraphics[angle =270, width=1.0\textwidth]{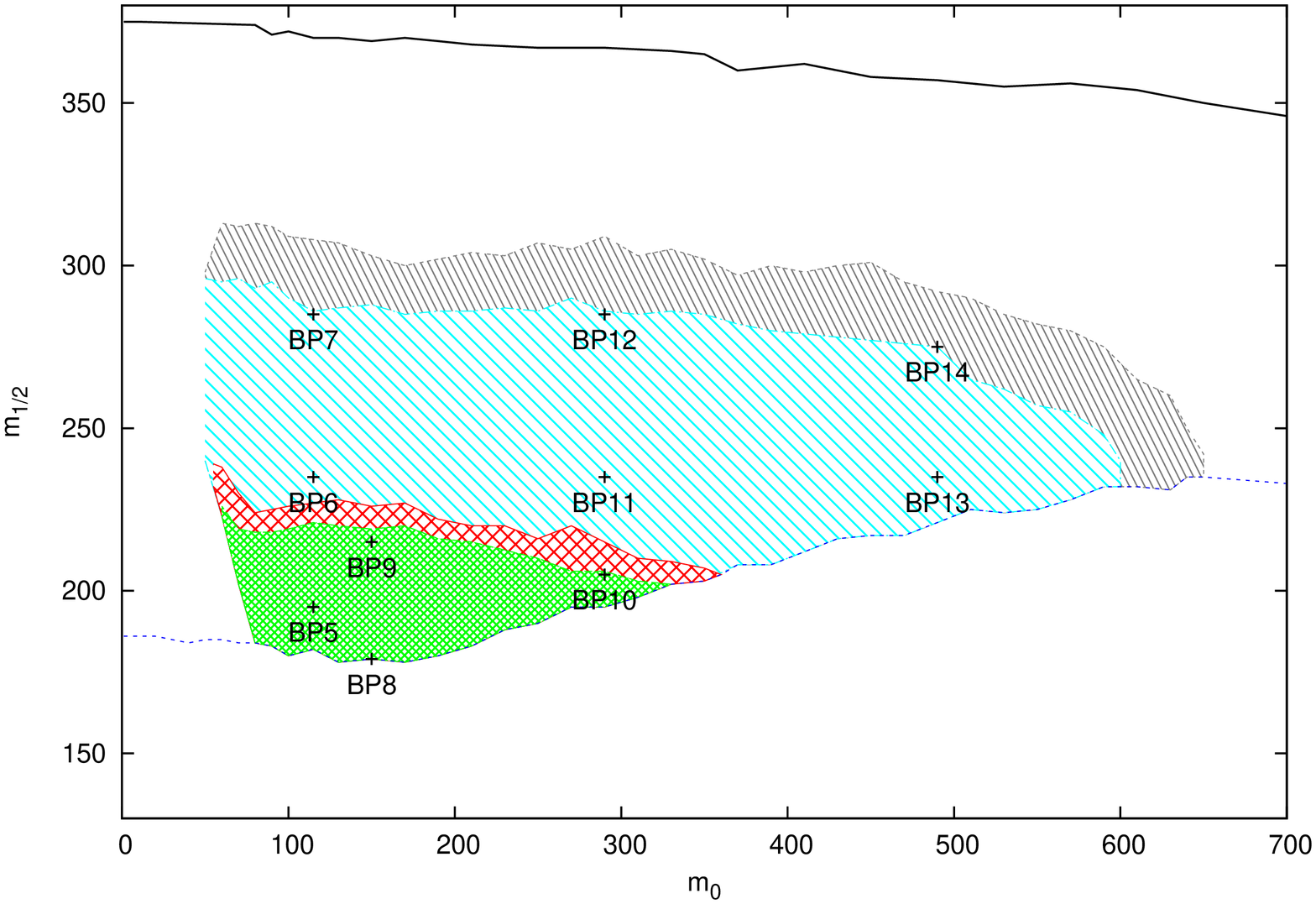}
\end{center}
\caption{Regions of $m_0$ - $m_{1/2}$ plane with $A_0 =-600$, 
tan$\beta =$ 5 and $sign(\mu) >$ 0 
which can be probed by the $1l + 1b + N_j(\ge 2) + \etslash$ signal
from all  squark-gluino events. The green (red) 
region gives observable signal $\lum =$ 1 (2) $\ifb$. 
The blue (grey) region is sensitive to the $1l + 1b + N_j(\ge 4) + \etslash$ 
signal from all squark-gluino events for $\lum =$ 1 (2) $\ifb$ (see Section 3.2).
The lower edge of the signal region is determined by the $m_h$$\approx$ 110 for
$A_0 =-600$ and tan$\beta =$ 5 (see Section 2 for details). The region bellow 
the upper black line corresponding to $m_h <110$  for $A_0 = 0$ and tan$\beta =$ 5 
and is in conflict with the $m_h$ bound from LEP.}

\end{figure}
\begin{table}[!htb]
\begin{center}\
 
\begin{tabular}{|c|c|c|c|c|c|}
\hline
\multicolumn{2}{|c|}{Points}      & \multicolumn{2}{c|}{$\lstop \lstop^*$} & \multicolumn{2}{c|}{$\tilde q$-$\tilde  g$(all)} \\
\hline
$m_0 $, $m_{1/2}$, $-A_0$ & $m_{\lstop}$ &  {\it Set 1} &  {\it Set 2}	&  {\it Set 1}	&  {\it Set 2}	\\
       \hline
130, 195,600	&175		&395	&313	&890	&393	\\
\hline
190,195,580	&205		&226	&183	&650	&236	\\
\hline
290,195,720	&191		&267	&215	&688	&257	\\
\hline
390,195,800	&218		&155	&128	&429	&149	\\
\hline
450,195,900	&216		&156	&127	&366	&152	\\
\hline
510,195,1020	&206		&208	&167	&364	&172	\\
    \hline

       \end{tabular}
       \end{center}
          \caption{ Separation of $\lstop \lstop^*$ events from all squark-gluino events 
for points selected from Fig. 3.}
          \end{table}

The missing energy spectrum of the signal from direct  $\lstop \lstop^*$  
production is weak (Fig. 2 (right)).
But $\tilde q$-$\tilde g$ events in general  will have high $\etslash$ 
and more number of jets. Thus, a new set of cuts  
has been designed to look for a more generalized signal
- $1l + 1b + N_j(\ge 4) + \etslash$.  

The following new cuts, collectively called {\it Set 3} are implemented 
for background rejection:
\begin{itemize}

\item We have selected events with one isolated lepton ({\it cut 1$^{\prime}$}).

\item We have selected events with one {\it tagged} $b$ jet ({\it cut 
2$^{\prime}$}).

\item Events with at least 4 central jets are selected, where central
jets are defined as pycell jets with $\vert \eta \vert \le$ 2.5 ({\it cut 
3$^{\prime}$}).

\item Events with missing transverse energy ($\etslash) \ge 200$   
are selected ({\it cut 4$^{\prime}$}).

\item Events with $M_{eff} \ge 600$   are selected ({\it cut 5$^{\prime}$}).

\end{itemize}

\begin{table}[!htb]
\begin{center}\
\begin{tabular}{|c|c|c|c|c||c|c|c|c|}
\hline
             & \multicolumn{4}{c|}{Signal} & \multicolumn{4}{c|}{Background} \\
\hline
              & BP1   & BP2  & BP3 & BP4       & $t \bar t$ &     QCD   & $W + 1j$    & $W + 2j$  \\ 
\hline
\hline

{\it cut 1$^{\prime}$} & 9.2115 &12.9063  & 2.5783 & 2.85    &28.728   &  2.2E+05     &3066.8  &953.33  \\ 
\hline   
{\it cut 2$^{\prime}$} &2.8417  &1.5424  & 1.0197  & 1.2017  & 14.0519 &  1.07E+04    & 8.2939 &  8.5925\\ 
\hline   
{\it cut 3$^{\prime}$} &1.0457  & 0.8610 & 0.7298  & 0.6243  & 4.6965  &   81.647     & 0.4349 & 1.4242 \\ 
\hline   
{\it cut 4$^{\prime}$} &  0.3399 &0.3296  & 0.2037  & 0.1553 & 0.0615  &   0.0016	& 0.0    & 0.0331 \\ 
\hline   
{\it cut 5$^{\prime}$} & 0.3106 & 0.3025 & 0.1963  & 0.1520  & 0.0487  &   0.0016	& 0.0    & 0.0189 \\ 
\hline
$S/\sqrt B$            &37.24   &36.34   & 23.58   & 18.26   &		&		&	 &	\\

\hline
\end{tabular}
\end{center}
\caption{Same as Table 6 but for cuts in {\it Set 3}.}
\end{table}


\begin{table}[!htb]
\begin{center}\
\begin{tabular}{|c|c|c|c|c|c|c|c|c|c|c|}
\hline
             & \multicolumn{10}{c|}{Signal}  \\
\hline
              & BP5    & BP6    & BP7   & BP8   & BP9   &  BP10  & BP11  & BP12 & BP13 & BP14   \\ 
$\sigma$ (pb) & 29.1   & 6.3    & 1.8   & 63.0  & 10.5  &  7.7   & 3.6   &  1.2 & 1.6  & 0.75   \\
\hline
\hline

{\it cut 1$^{\prime}$}    &8.7852	&1.5774	&0.4144	&21.6846& 3.03	&2.1667	&0.9301	&0.2731	&0.5269	&0.2352	\\
\hline   
{\it cut 2$^{\prime}$}    &2.3066	&0.5119	&0.1103	&2.1545	&1.1994	&0.9003	&0.3688	&0.0968	&0.1915	&0.0859	\\
\hline   
{\it cut 3${^\prime}$}    &0.7954	&0.2484	&0.0727	& 1.1823& 0.5641&0.5993	&0.2815	&0.0768	&0.1765	&0.0771	\\
\hline   
{\it cut 4$^{\prime}$}    &0.2987	&0.1165	&0.0436	&0.4032	&0.2019	&0.2127	&0.1222	&0.0426	&0.0705	&0.0423	\\
\hline   
{\it cut 5$^{\prime}$}    &0.2793	&0.1129	&0.0414	&0.3612	&0.1911	&0.2048	&0.1184	&0.0418	&0.0693	&0.0418	\\
\hline 
$S/\sqrt B$               & 33.56       & 13.56 & 5.0   & 43.4  & 22.96 & 24.61 & 14.23 & 5.02  & 8.33  & 5.02  \\

\hline
\end{tabular}
\end{center}
\caption{Same as Table 8 but for BP5-14 for cuts in {\it Set 3}.}
\end{table}


All possible SM backgrounds as listed in the previous section have been
computed with these revised cuts. In Table 8 and Table 9 the efficiencies
of the {\it Set 3} of cuts for the signal and SM background are listed.
The significance of the signal for $\lum =$ 1 $\ifb$ is given in the last row of 
Table 8 and 9 for different benchmark points.
In Fig. 4 the blue (grey) region yield observable signals 
in this channel for $\lum =$ 1 $\ifb$ (2 $\ifb$). The relative strong cuts, 
however, eliminate a large fraction of the events from direct  $\lstop \lstop^*$ 
production and reconstruction of $\lstop$ mass is not possible.

Finally to compare the $1l + 1b + N_j(\ge 2) + \etslash$ signal
and the $1l + 1b + N_j(\ge 4) + \etslash$ signal with the canonical 
$Jets + \etslash$ signal, we  compute the latter signal using  the cuts from \cite{atlas}.
The selection criteria, hereby called {\it Set 4} are as follows :

\begin{itemize}

\item Events with isolated leptons with $P_T \ge 10$   are rejected. 

\item Events are selected with $N_j \ge 2$.

\item We further demand events with $P^{j1}_T \ge 70$   and all other jets
with $P_T \ge 30$.
\item Events with $\etslash \ge 40$   are selected.
\item We select events with $\Delta \phi(\etslash,j_i) > 0.2$.
\item The selected events must have the ratio of $\etslash$ and $M_{eff}$ 
greater than 0.3.

\end{itemize}

\begin{table}[tb]
\begin{center}\
\begin{tabular}{|c|c|c||c|c||c|c|}
\hline
             & \multicolumn{2}{c|}{{\it Set 1}} & \multicolumn{2}{c|}{
{\it Set 3 }} & \multicolumn{2}{c|}{ {\it Set 4}} \\
\hline
 Points      &$	A_0 \neq 0$ & $	A_0 = 0$ &$A_0 \neq 0$ & $A_0 = 0$ & $	A_0 \neq 0$ & $	A_0 = 0$ \\ 
\hline
\hline
BP1		&890	&190	&310	&114	&4175	&4179	\\ 
BP2		&594	&187	&302	&114	&4392	&3957	\\ 
BP3		&346	&108	&196	&97	&792	&939	\\ 
BP4		&366	&70	&152	&59	&545	&366	\\ 
BP5		&811	&181	&279	&67	&5039	&4254	\\ 
BP6		&165	&67	&113	&101	&1740	&1564	\\ 
BP7		&35	&19	&41	&26	&563	&510	\\ 
BP8		&844	&306	&361	&162	& 6806	& 6109	\\ 
BP9		&396	&109	&191	&86	&2195	&2376	\\ 
BP10		&321	&162	&204	&137	&1312	&1398	\\ 
BP11		&132	&88	&118	&87	&671	& 782	\\ 
BP12		&31	&26	&42	&36	&305	&314	\\ 
BP13		&66	&23	&69	&24	&103	&157	\\ 
BP14		&28	&18	&42	&24	&84	& 91	\\ 
\hline
\end{tabular}
\end{center}
\caption{ The number of squark-gluino events in three different channels for 
$\lum = 1 \ifb$ with $A_0 \ne 0$ and $A_0 =0 $ : i) the channel $1l + 1b + N_j(\ge 2) + \etslash$ 
using the cuts in {\it Set 1}(columns 2 and 3), ii)the channel $1l + 1b + N_j(\ge 4) + \etslash$ 
using the cuts in {\it Set 3}(columns 4 and 5) and iii)the $jets + \etslash$ channel 
using the cuts in {\it Set 4}(columns 6 and 7).}
\end{table}

In Table 10, the size of the squark-gluino signal for {\it Set 1}, {\it 
Set 3} and {\it Set 4} cuts for $\lum =$ 1 $\ifb$ corresponding to 
different BPs are given for $A_0 = 0$ and $A_0 \ne 0$ . Comparing column 
2 and column 3 we indeed find that the signal size is strikingly 
different for the two cases . The same conclusion follow by comparing 
columns 4 and 5 . One can also see this difference by observing the 
ratio of signal size with the $Jets + \etslash$ signal for $A_0 \ne 0$ 
and $A_0 = 0$ (columns 6 and 7). It may be recalled that this ratio is 
fairly insensitive to theoretical uncertainties like the choice of the 
QCD scale, the choice of PDFs etc.

In  \cite{debottam} it was emphasized that regions of the mSUGRA parameter 
space corresponding to low $m_0$ and $m_{1/2}$   
are compatible with the  WMAP data \cite{wmap} on the DM relic density
of the universe and the $m_h$ bound only for $A_0 \ne 0$. 
In Fig.4 the lower edge of the signal region corresponds to the 
$m_h$ = 110 line for tan$\beta =$ 5 and $A_0$ = -600. The line above 
the signal region represents the same line for $A_0$ = 0. 

In the same figure the region consistent with the WMAP data can be 
identified by comparing with Fig 2 of \cite{debottam}. We find that
for $m_0 = 80$, tan$\beta =$ 5 and $A_0$ = -600, the region 320$\lsim 
\mhalf \lsim$ 400 is consistent with the WMAP data. The corresponding 
region for $m_0$ = 120 is 470$\lsim \mhalf \lsim$ 550. 
Apparently only a small fraction of the signal
region near and just above the left upper edge is consistent with the
WMAP data. However, it should be borne in mind that our signal estimates
are based on the leading order cross sections. As discussed above more 
accurate 
estimates based on the NLO cross sections and optimized cuts are 
expected to yield better reach. Integrated luminosity higher than 2
$\ifb$ may  improve the  reach further.
        
A larger parameter space consistent with the WMAP data which yields
the signal can be found if the $A_0 $  and tan$\beta$  are allowed 
to vary.
In \cite{debottam} the region of the mSUGRA parameter space consistent
with WMAP data for
other choices of the above two parameters were also identified.
In Table 11 we present
several such points. The corresponding signals are computed 
subject to the cuts of {\it Set 3}.
We also present the significance of each signal for $\lum =$ 1 $\ifb$. 
It should be noted that observable signals are possible for 
moderate values of $A_0$ and higher values of tan$\beta$.

\begin{table}[!htb]
\begin{center}\
\begin{tabular}{|c|c|}
\hline
$m_0$, $m_{1/2}$, $-A_0$, tan$\beta$    &                $S/\sqrt B$ \\
\hline
80,350,700,5       &                    5.4 \\
80,300,1000,5      &                    5.6 \\
100,200,700,10     &                    25.2 \\
100,220,700,10     &                    16.6 \\
120,280,900,10     &                    5.16 \\
150,200,600,20     &                    24.2 \\
170,200,300,30     &                    23.1  \\
170,220,350,30     &                    15.0 \\
\hline
\end{tabular}
\end{center}
\caption{Significance of the squark-gluino signals corresponding to different mSUGRA points
taken from \cite{debottam} compatible with the WMAP data. For all
points sign$(\mu) > $ 0.}
\end{table}
\section {Conclusions}

LHC experiments at the early stage of the 7 TeV run are sensitive to the 
mSUGRA parameter space with relatively low $m_0 - m_{1/2}$ . A sizable part 
of this parameter space is excluded for low values of $A_0$ and 
tan$\beta$ by the bound $m_{h} > $ 114.4 from LEP. If mSUGRA is realized 
in nature with rather low sparticle masses it is , therefore, likely 
that either $A_0$ or tan$\beta$ or both should be large. Additional 
interest in this region stems from the fact that here LSP bulk 
annihilation and/or stau-LSP co-annihilation may produce the observed 
dark matter relic density. If 
SUSY is discovered during the early runs at 7 TeV, the validity of 
mSUGRA can be tested if some additional information on $A_0$ and tan
$\beta$ is available.

The canonical $Jets + \etslash$ signature, however, is not very 
sensitive to $A_0$ . In this paper we suggest the signature 
$(blj$$\etslash$) which is sensitive to moderate and large values of 
$A_0$ and low tan$\beta$ and can complement the canonical signal. This 
signal may arise either from direct $\lstop \lstop^* $ pair production 
or from $\tilde g \ra \tilde t_{1} t $ provided $\lstop$ is much lighter 
than all other squarks. This naturally happens in models with low 
tan$\beta$ and large $A_0$.

In Section 2 we have motivated this signature by introducing several
benchmark points. Some of the chosen points have features which naturally 
ensure a strong signal. Some, on the other hand, are selected such that one or
more of the above features are absent, leading to relatively weak signals. Next we scan over 
the parameter space containing these points and analyze the visibility
of the signal during 7 TeV run.

In Section 3 we examine the proposed signals. In 3.1 the $blj$$\etslash$ 
signal with $N_j \ge2$ from direct 
$\lstop \lstop^*$ pair production has been studied with the assumption that 
all other strongly interacting sparticles are beyond the reach of the ongoing 
run at low luminosities. Using the {\it Set 1} of cuts (Section 3.1), we estimate that 
$m_{\lstop} <$ 192(220)   can be probed with $\lum = $ 1$\ifb$ (5$\ifb$) (See Fig. 3).
We stress that these estimates based on the LO cross-sections are conservative. 
Better mass reach is likely to follow by using the NLO cross-sections as discussed 
in the text. As pointed out at the end of the Section 3.1, certain departures
from the sparticle spectrum of mSUGRA may make the $\lstop $ search prospect 
even better.

A much larger region of the mSUGRA parameter space can be scanned by analyzing all
squark-gluino events using the {\it Set 1} of cuts (Section 3.2). This is illustrated in Fig. 3 and 4 (see also Tables 6).

In order to study the properties of $\lstop $ we suggest a procedure for filtering
out the $\lstop \lstop^*$ events from all SUSY events using the {\it Set 2} of cuts 
introduced in Section 3.2 . We find that nearly 80$\%$ pure $\lstop \lstop^*$ sample 
can be separated by this method (see Table 7). It will be interesting to see 
whether $m_{\lstop} $ can be reconstructed by using the standard procedures 
\cite{mt2} and this will be applicable for higher sparticle masses within 
the reach of LHC 14 TeV runs. Our results, however, indicate that the filtering 
could be more efficient for heavier squarks and gluinos.

The directly observable $\lstop \lstop^*$ events has a rather soft $\etslash$ 
spectrum. Thus a strong $\etslash $ cut - a very potent tool for suppressing 
the SM background - can not be fully utilized. We have shown that by using stronger
cuts ({\it Set 3}) a much larger region of the $ m_0$-$m_{1/2}$ space can be probed via 
all the squark-gluino events and the presence of the underlying non-zero trilinear coupling
can be traced. Using the {\it Set 3} of cuts in Section 3.2 the parameter space 
delineated in Fig. 4 can be probed by the 
$1l + 1b + N_j(\ge 4) + \etslash$  signal (see also Table 8-9). It is interesting 
to note that several  parts of the parameter space with
$A_0 \ne 0$
yield relic densities consistent with WMAP data as well as observable
signals (see  Table 11).

Finally in Table 10 we compare the $1l + 1b + N_j(\ge 2) + \etslash$ signal using the {\it Set 1} 
of cuts for vanishing and non-vanishing trilinear couplings (see columns 2 and 3).
The same comparison for the $1l + 1b + N_j(\ge 4) + \etslash$  
using the {\it Set 3} of cuts is also presented in the same table (see columns 4 and 5). 
It is clear that both the signals are rather sensitive to the value of 
$A_0$. In the last columns of this table we present the canonical $Jets +\etslash$ 
signal for the above two choices of $A_0$. We find that - as expected - the signal 
is more or less insensitive to $A_0$. More importantly the ratio 
of various observables in Table 10 are insensitive to a 
large extent to the theoretical uncertainties like the choices of the QCD scale.

We believe that if SUSY is indeed discovered in the $Jets +\etslash$ channel, the 
signal proposed in this paper may provide complementary information about the 
trilinear coupling in the mSUGRA model for both 
7 TeV and 14 TeV runs.

\vspace{0.5 cm}
{\Large \bf Note Added :}

When our work was in the final stage, the CMS collaboration published 
the first result of SUSY search in the $Jets +\etslash$ channel at the 
ongoing LHC experiments for 35 ${pb}^{-1}$ of data \cite{CMS1}. The 
analysis was done in the mSUGRA model. The main result is an exclusion 
plot in the $m_0$-$m_{1/2}$ plane for $A_0 = 0$ and tan$\beta$ = 3 and 
$sign(\mu)> 0$. From the discussions in the introduction it is clear 
that this excluded region is incompatible with the $m_h$ bound. 

It may, however, be argued that the $Jets +\etslash$ signature is fairly 
insensitive to $A_0$ and tan$\beta$ \footnote{The last two columns of 
Table 10 support the observation. For a recent illustration of this 
point see the analysis of the $Jets +\etslash$ data by the ATLAS 
Collaboration \cite{ATLAS2}. See also Fig. 1 (right) in \cite{prannath}. 
A part of this paper generalizes the ATLAS exclusion plots for large 
tan$\beta$ and $A_0$.}. With this approximation the parameter spaces 
with $A_0 \ne 0$ sensitive to our signals can be examined in the light 
of the CMS exclusion plot.  

Most of the parameter space sensitive to the 
$1b+1l+N_j(\ge2)+\etslash$ signal arising from all squark-gluino 
production (see the green and red regions of Fig. 3 and 
Fig. 4) is disfavoured by the CMS exclusion plot. However, the 
separation of the stop signal from squark-gluino events as 
discussed here could be useful in future.  
Finally a large region sensitive to the $1b+1l+N_j(\ge4)+\etslash$ (see 
blue and grey regions of Fig. 4) survives the CMS exclusion 
plot.

Subsequently the ATLAS collaboration published a similar exclusion plot 
using the $1l+jets+\etslash$ data based on the same set of mSUGRA 
parameters disfavoured by the $m_h$ bound \cite{ATLAS1}. Apparently 
their observed limits are stronger than the corresponding CMS 
exclusion. This result is very different from the expectations based on 
earlier experiments or simulations. A closer scrutiny, however, reveals 
that they have observed 1 event each for the $e$ and the $\mu$ channels 
against backgrounds of 1.81$\pm$ 0.75 and 2.25 $\pm$ 0.94 respectively. 
This signal deficit which could be due to fluctuations in a low 
statistics experiment is one of the reasons for their stronger limits 
\footnote { We thank Dr Satyaki Bhattacharya for a very fruitful 
discussion on this point. }. We are of the opinion that at this stage of 
the LHC experiment emphasis should be given on the expected limit or the 
median expected limit of \cite{ATLAS1} both of which yield much weaker 
exclusion almost similar to the CMS result \cite{CMS1}. 

As discussed in the introduction, the signals involving 
$e$ or $\mu$ are very sensitive to tan$\beta$ and is strictly valid for 
low tan$\beta$ only. This point 
has recently been emphasized in \cite{debottam,adnabanita}. 

However the 
very recent ATLAS exclusion plot using the $jets +\etslash$ data 
\cite{ATLAS2} is somewhat stronger than the corresponding CMS plot. This 
will exclude more blue and grey regions sensitive to the 
$1b+1l+N_j(\ge4)+\etslash$ signal in Fig. 4. Nevertheless a sizable 
parameter space, the large $m_0$ part of it in particular, survives this 
exclusion.

{\bf Acknowledgment}:

NB acknowledge the Council of Scientific and Industrial Research (CSIR),
India for a research fellowship. We would like to thank Dr. Biplob
Bhattacherjee for help in computation.



\end{document}